\pdfoutput=1
\documentclass[british,mathic=true,xypdf,longnamesfirst]{scrartcl}
\usepackage[T1]{fontenc}
\usepackage{color}
\usepackage{babel}
\usepackage{prettyref}
\usepackage{float}
\usepackage{booktabs}
\usepackage{mathtools}
\usepackage{enumitem}
\usepackage{varwidth}
\usepackage{amsmath}
\usepackage{amsthm}
\usepackage[authoryear]{natbib}
\PassOptionsToPackage{hyphens}{url}
\usepackage[unicode=true,
 bookmarks=true,bookmarksnumbered=false,bookmarksopen=false,
 breaklinks=true,pdfborder={0 0 0},pdfborderstyle={},backref=page,colorlinks=true]
 {hyperref}
\hypersetup{pdftitle={Resource Polymorphism},
 pdfauthor={Guillaume Munch-Maccagnoni},
 ocgcolorlinks,linkcolor={dark-red},citecolor={dark-blue},urlcolor={blue}}
\makeatletter
\providecommand{\tabularnewline}{\\}
\PassOptionsToPackage{caption=false}{subfig}%
\usepackage{adjustbox}
\usepackage{xcolor}
  \definecolor{dark-red}{rgb}{0.6,0,0}
  \definecolor{dark-blue}{rgb}{0,0.0,0.6}
  \definecolor{blue}{rgb}{0,0,0.8}

\usepackage[expansion=false]{microtype}

\def\norsfs{\@namedef{ver@mathrsfs.sty}{9999/12/31}}
\norsfs
\ifxetex
  \hypersetup{unicode,psdextra}
  \PassOptionsToPackage{fixxits}{perfectcut}
  \usepackage{fontspec}
  \defaultfontfeatures{Ligatures=TeX}
  \usepackage[math-style=TeX]{unicode-math}
  \ExplSyntaxOn
  \AtBeginDocument{\cs_set_eq:NN \not \not_newnot:N}
  \ExplSyntaxOff
\else%
  \usepackage{stix} %
  \usepackage[scaled=0.84]{helvet}
  \usepackage[scaled=0.95]{inconsolata}%
  \usepackage[scr=boondoxo,cal=pxtx%
                             ]{mathalfa}
\fi%

\usepackage{etex}
\usepackage{bussproofs}
\usepackage{mleftright}
\mleftright
\hypersetup{unicode,psdextra}
\proofSkipAmount{\vskip1ex plus.8ex minus.4ex}

\usepackage{multirow}
\g@addto@macro\bfseries{\boldmath}
\usepackage{enumitem}
\def\mynobreakpar{\@beginparpenalty=10000} 
\usepackage{flushend}
\newrefformat{prop}{Proposition \ref{#1}}
\newrefformat{subsec}{Section \ref{#1}}
\newrefformat{rem}{Remark \ref{#1}}
\newrefformat{cor}{Corollary \ref{#1}}
\usepackage{listings}
\makeatother

\begin{document}
\title{Resource Polymorphism}
\author{Guillaume Munch-Maccagnoni (Inria, LS2N CNRS)\\
\href{mailto:Guillaume.Munch-Maccagnoni@Inria.fr}{Guillaume.Munch-Maccagnoni@Inria.fr}}
\date{6th March 2018}
\maketitle
\begin{abstract}
We present a resource-management model for ML-style programming languages,
designed to be compatible with the OCaml philosophy and runtime model.
This is a proposal to extend the OCaml language with destructors,
move semantics, and resource polymorphism, to improve its safety,
efficiency, interoperability, and expressiveness. It builds on the
ownership-and-borrowing models of systems programming languages (Cyclone,
C++11, Rust) and on linear types in functional programming (Linear
Lisp, Clean, Alms). It continues a synthesis of resources from systems
programming and resources in linear logic initiated by Baker.

It is a combination of many known and some new ideas. On the novel
side, it highlights the good mathematical structure of Stroustrup's
\emph{``Resource acquisition is initialisation\textquotedblright{}}
(RAII) idiom for resource management based on destructors, a notion
sometimes confused with finalizers, and builds on it a notion of resource
polymorphism, inspired by polarisation in proof theory, that mixes
C++'s RAII and a tracing garbage collector (GC). In particular, it
proposes to identify the types of GCed values with types with trivial
destructor: from this definition it deduces a model in which GC is
the default allocation mode, and where GCed values can be used without
restriction both in owning and borrowing contexts.

The proposal targets a new spot in the design space, with an automatic
and predictable resource-management model, at the same time based
on lightweight and expressive language abstractions. It is backwards-compatible:
current code is expected to run with the same performance, the new
abstractions fully combine with the current ones, and it supports
a resource-polymorphic extension of libraries. It does so with only
a few additions to the runtime, and it integrates with the current
GC implementation. It is also compatible with the upcoming multicore
extension, and suggests that the Rust model for eliminating data-races
applies.

Interesting questions arise for a safe and practical type system,
many of which have already been thoroughly investigated in the languages
and prototypes Cyclone, Rust, and Alms.
\end{abstract}

\clearpage{}

\tableofcontents{}

\section{Introduction}

A resource is a value that is hard to copy or dispose of. A typical
resource is a large data structure, a file handle, a socket, a lock,
a value from a cache, an exclusive access to a mutable, a value originating
from a different runtime, or a continuation. It is also any data structure
composed from such resources, such as a list of resources or a closure
containing a resource. Support for resource management in programming
languages (PLs) is a concern for safety, efficiency, interoperability
and expressiveness.

This is a proposal for a resource-management model compatible in broad
strokes with the OCaml\footnote{\href{https://ocaml.org/}{https://ocaml.org/}}
philosophy and runtime model. By abstracting a few low-level technical
details, it can also be read more generally as a model for other languages
in the ML family.

It considers new values and types that own or borrow resources, similar
to ownership/borrowing in C++11\footnote{\href{https://isocpp.org/}{https://isocpp.org/}}/Rust\footnote{\href{https://www.rust-lang.org/}{https://www.rust-lang.org/}},
in addition to the current GC types. It is motivated by addressing
issues that arose during discussions with several Serious Industrial
OCaml Users concerning safety, efficiency, interoperability, and expressiveness
in the presence of resources: the inadequacy of finalizers for timely
disposal of large pools of resources, the unpredictability and limitations
of unboxing, the difficulty of interfacing with resource-intensive
libraries such as Qt\footnote{\href{https://www.qt.io/}{https://www.qt.io/}},
the difficulty of cleaning-up resources reliably with fibers in ocaml-multicore\footnote{\href{https://github.com/ocamllabs/ocaml-multicore}{https://github.com/ocamllabs/ocaml-multicore}},
the need for affine closures with effect handlers, and more.

By OCaml philosophy, what is meant is reaching a sweet spot combining:
\begin{enumerate}
\item A safe type system that helps instead of hindering,
\item Lightweight and expressive abstractions,
\item An efficient runtime.
\end{enumerate}
This proposal focuses on the levels 2 and 3 above. It is voluntarily
vague about level 1, for the reason that resource-friendliness is
deeply rooted in the computational behaviour (i.e. level 3), as it
will become clear. Let us put aside the idea that one can start from
ideas for a type system and expect interesting computational behaviour
to suddenly appear; instead, the type system should come in a second
time, in service of a convincing design for computational aspects.
The first challenge for OCaml, tackled here, is to get level 2 right,
such that level 3 can become a realistic, conservative, and useful
extension of the current runtime. Level 1 is expected to require substantial
effort; hopefully the proposal provides sufficient motivations for
such an effort. Besides, we will see that there is ample prior work
addressing various questions for a practical type system, already
at work separately in the languages Rust and Alms.

The model is inspired by Stroustrup's RAII \citep{Stroustrup1994}
and Hinnant et al.'s move semantics in C++11 \citep*{Hinnant2002}.
RAII (\emph{\textquotedblleft Resource acquisition is initialisation\textquotedblright })
proposes to integrate error handling and resource management by attaching
destructors to types: clean-up functions that are called automatically
and predictably when a scope ends, whether by returning or due to
an exception being raised. It is an essential ingredient in the \emph{basic
exception-safety guarantee} \citep{Stroustrup2001} which requires
that functions that raise an exception do not leak resources and leave
all data in a valid state. In the words of \citet*{Ramananandro2012},
RAII enforces invariants about the construction and the destruction
of resources predictably and reliably.

In contrast, \emph{finalizers} as currently used in OCaml, that is
clean-up functions called by the garbage collector, are not predictable,
are not guaranteed to be run, and allow making values reachable again
(\citealp*[Chapter 21]{Minsky2013}; similar points are made for finalizers
in other languages in \citealp*{Stroustrup2015}). In OCaml, which
thread calls the finalizer is even explicitly unspecified\footnote{\href{http://caml.inria.fr/pub/docs/manual-ocaml/libref/Gc.html}{http://caml.inria.fr/pub/docs/manual-ocaml/libref/Gc.html}}.
Finalizers appear commonly considered inappropriate for managing resources.

In addition, while the original resource-management model of C++ was
criticised for its over-reliance on deeply copying values (among others),
Hinnant et al.\ proposed to introduce a new kind of types in C++
(\emph{rvalue references}) that allowed to express the moving of resources.
In particular:
\begin{itemize}
\item it supported a movable and non-copiable pointer for automatic resource
management (\emph{uni\-que\_ptr}), reminiscent of uniqueness \citep{Barendsen1996}
and ownership types \citep{Clarke1998},
\item it supported a polymorphism of resource management: the management
of a type is by default deduced from its components,
\item it supported the conservative extension of data structures and algorithms
from the standard library to operate on resources,
\item all the while retaining backwards compatibility with existing code
(sometimes even speeding it up by removing unnecessary copies).
\end{itemize}
Together with the extensive use of (unsafe) passing by reference and
a rudimentary form of reference-counting garbage collection expressible
with RAII (\emph{shared\_ptr}), this is advocated as the new resource-management
model of C++11 (\citealp*{Stroustrup2015}). Ownership types, and
regions as in MLKit \citep{Tofte1998} and Cyclone \citep{Jim2002,Grossman2002},
have also been proposed as abstractions amenable to static analyses,
which has inspired Rust's ownership-and-borrowing model strengthening
the C++11 model with static safety guarantees and a novel design for
preventing data races \citep{Anderson2016}.

This proposal, however, should not be seen as just trying to extend
OCaml with C++ idioms. Its starting point was the similarities between
C++'s resource polymorphism and polarisation in proof theory, as well
as a rational reconstruction of destructors in the linear call-by-push-value
categorical model \citep*{CFM2015}. This suggested several aspects
of this proposal, by bringing to light the deep compatibility of the
C++11/Rust resource-management model with functional programming.
This continues a synthesis of systems programming's resources and
linear logic's resources initiated in a series of rarely-mentioned
articles by \citet{Baker1994,Baker1994a,Baker1995}.

In the end, polishing RAII brings additional similarities with Rust,
a runtime model that fits that of OCaml, and applications that go
beyond a simple replacement of finalizers. This proposal is an element
in a broader thesis that RAII hides a fundamental computational structure
that has not been given yet the exposure it deserves.

\subsection{\label{subsec:The-relevance-of}The relevance of C++11 for OCaml}

The proposal will remind of linear types, regions, uniqueness types,
ownership types, and borrowing, \emph{à la} Linear Lisp, Clean, Cyclone,
Rust, etc. (\prettyref{sec:Comparison-with-existing} offers a more
detailed comparison.) But in comparison to Linear Lisp, Clean, Cyclone,
Rust, etc., the experience of the move from C++98 to C++11 stands
out for OCaml for three reasons:
\begin{itemize}
\item Both are established languages that need to preserve the meaning and
performance of large amounts of legacy code.
\item Both have to deal with exceptions, a core part of their design, much
more prominently than in Rust in which exceptions (\emph{panic}) are
restrictive and discouraged by design, or than in other languages
in which they are absent.
\item Both are designed around light, efficient, and predictable abstractions.
\end{itemize}
Of course, C++11 and OCaml differ greatly in other technical and principal
aspects, and their communities do not overlap much, which would explain
why, if there is any value to this proposal, it has not been proposed
before, besides the example set by Rust. In addition, some experience
with mathematical models of PLs in the categorical tradition has helped
(the author at least) reading between the lines of the C++11 specification
and of various idioms that arose from it, and extracting its substance.
The inputs from semantics are explained in the next section.

While the crucial ideas for this proposal are RAII and move semantics
from Stroustrup and Hinnant et al., the end result is probably closer
to Rust. This is because Rust itself was inspired by both C++ and
ML among others \citep{Anderson2016}. Compared to C++, Rust offered
language support for isolating the \emph{\textquotedblleft unsafe\textquotedblright{}}
parts of the code in libraries. This means that in most of user code,
what are merely \emph{best practices} in C++11 are enforced by the
Rust compiler, thereby providing strong static safety guarantees.
Moreover, by treating mutable state as a resource, Rust tracks aliasing,
ensuring that no data races are due to bugs in \emph{\textquotedblleft safe\textquotedblright{}}
mode. This tour de force for an industrial language drew the attention
of the academic PL community. C++ also takes example on Rust with
the ongoing \emph{Core Guidelines initiative} \citep{Stroustrup2015Core}
aimed at standardizing and tooling a \emph{\textquotedblleft smaller,
simpler, safer language\textquotedblright }, with similar ideas as
Rust but an emphasis on easy migration of legacy code.

This proposal suggests that a similar path is possible for OCaml,
where \emph{\textquotedblleft small, simple\textquotedblright{}} (and
efficient) is retained from the OCaml that everyone likes, and where
\emph{\textquotedblleft safer\textquotedblright{}} is achieved for
resources and concurrency, without sacrificing the expressiveness
of a functional, GCed language. This proposal focuses on resource
safety; Rust also tries to solve the problem of data races. In the
current proposal, OCaml's unrestricted shared mutable state is kept
for the sake of backwards compatibility. Proposing a solution to data
races in OCaml \emph{à la }Rust in one go would be ambitious. The
requirement of backwards compatibility can seem a convenient excuse
to not propose one right away, but in fact it shows an opportunity
to first integrate some language features that are already useful
for resource management, while at the same time providing a richer
playground for tackling concurrency problems in the future. Nevertheless,
\prettyref{subsec:Languages-with-control} comes back on this suggestion
on an optimistic note.

\subsection{\label{subsec:RAII,-resource-polymorphism,}RAII, resource polymorphism,
and GC, from a semantic point of view}

RAII is an idiom in which a destructor is attached to a resource,
according to its type. The destructor is called predictably when the
variable goes out of scope, including due to an exception being raised.
With RAII, one can allocate resources on the heap, and have the resources
automatically collected predictably and reliably, bypassing garbage
collection. The destructor deallocates memory, and can be customised
by the language or the user to perform other clean-up duties.

Three intuitions arising from mathematical models were the starting
points of this proposal:
\begin{enumerate}
\item Hinnant et al.'s resource polymorphism for C++11 coincides with Girard's
polarity tables in proof theory, which describes compound connectives
in terms of basic ones \citep{Gir91,Gir93}. In proof theory, the
goal of polarity tables is to minimize the number of modalities, so
as to maximize the number of type isomorphisms \citep{Gir91} and
``$\eta$'' conversions \citep*{danos95new}; for PLs these concerns
have an immediate application in making programs easier to reason
about. (\citealt{Gir93} describes the mixing of linear, intuitionistic,
and classical logics in the same system, which in fact corresponds
to a mixing of resources and continuations.) For the purposes of this
proposal, a polarity is a type of types closed under constructions.
This view is close to that of \emph{\textquotedblleft kinds as calling
convention\textquotedblright{}} \citep{Bolingbroke2009,Eisenberg2017},
itself inspired in part by Levy (\citeauthor{Bolingbroke2009}).

Both Girard and Hinnant et al.\ describe systems in which, instead
of a non-copyable pair $\otimes$ distinct from a copyable one (maybe
written differently, $\times$), there is a single pair $\otimes$,
whose polarity (e.g. whether it is copyable or not) is deduced from
its components: $A\otimes B$ is copyable by default as soon as both
$A$ and $B$ are; in this case the copy operation consists in the
sequence of copies of members, which turns out to have an abstract
description in the form of a canonical algebraic structure \citep{Bierman}.
(With linearity interpreted as counting uses, a similar idea was proposed
in some form by the Clean language, and it has reappeared several
times. See \prettyref{subsec:Aside-Origins-of}.)
\item We have noticed, in joint work with Guillaume Combette, that the notion
of scoped-tied destructors arises naturally when modelling exceptions
and local control (return, break...) in the linear call-by-push-value
(LCBPV) model of effects and resources. CBPV \citep{Levy99CBPV,Levy2004}
can be seen as an idealised model of ML-style languages (higher-order,
typed, with strict data types and effects) refining the call-by-value
$\lambda$-calculus, and LCBPV is a natural decomposition of it generalizing
linear logic \citep{Gir87} with hopes that it could serve to model
the interaction of effects and resources. Nothing prefigured that
the model had to do with idioms from a systems programming language.

The lesson is that trying to model exceptions or local control in
LCBPV naturally leads to the rediscovery of scope-based destructors,
and of several of their peculiarities which are described below. It
provides the perspective that the interpretation of resources as affine
types is not at odds with linear logic, but in fact arises from it.
In contrast, LCBPV does not justify other resource-management idioms:
it attributes no meaning to finalizers, and \emph{try...finally} (e.g.
Java) is as \emph{ad-hoc} as one expects. This suggests that RAII
is a fundamental computing concept, similarly to CPS given that it
arises from the same kind of algebraic considerations.
\item While the previous points suggest that a practical resource-management
model has to mix different resource-management techniques as determined
by the type, it still remains to explain how an ownership-and-borrowing
model can be integrated in a language with a GC. For this, let us
draw from the consensus that the GC should not have to perform non-trivial
finalization, and make it a definitional principle:
\begin{quote}
the GC is a run-time optimisation that either delays or anticipates
the collection of values that can be trivially disposed of.
\end{quote}
From this definition, we will derive a way of mixing GCed values with
resources, alternative to using finalizers.
\end{enumerate}
In any case, there is a leap of faith between the linear call-by-push-value
models as they currently stand and the current proposal. It would
be pointless to get into more mathematical details at this point.
Then, what is the value of the abstract point of view? To begin with,
a contribution of the semantic view on RAII is to reassure about various
peculiarities which might otherwise seem \emph{ad-hoc}:
\begin{itemize}
\item It gives rise to a notion of affine types that, instead of looking
at odds with the linear logic narrative (why affine rather than linear?),
arises naturally from it, and happens to match idioms used in successful
industrial languages.
\item It sheds a natural light on complicated rules, such as the rules for
automatic generation of destructors (for instance in C++ if two types
$A$ and $B$ have destructors $d_{A}$ and $d_{B}$ then $A\otimes B$
has destructor $d_{A}\otimes d_{B}$ which performs $d_{A}$ and $d_{B}$
in sequence). Such rules actually describe a canonical mathematical
construction and therefore enjoy good properties not necessarily visible
from the surface of a language.
\item It attributes no meaning to exceptions escaping from destructors.
In C++ too, this is undefined behaviour: otherwise one can end up
with several exceptions being raised at the same time if destructors
throw during stack unwinding.
\item It explains peculiarities of pattern-matching in the presence of ownership:
the common explanation in terms of universal properties seems to recover
the intuitive fact that assuming ownership during pattern matching
is only possible if the destructor of the pattern is the default one.
This predicts an integration of ownership with pattern matching already
explored by Rust.
\end{itemize}
The value of the model is therefore to encourage a bold change, to
which one would not necessarily come in small steps by trying to fix
separately the various issues related to resource management in OCaml.
We hope to explain the (modest) mathematical aspects in greater detail
elsewhere.

\subsection*{Outline}

\prettyref{sec:Integrating-ownership-and} describes the integration
of ownership and borrowing with a traced garbage collector based on
a notion of resource polymorphism. \prettyref{subsec:Aside-Origins-of}
come back on move semantics and resource polymorphism from a historical
perspective. From \prettyref{sec:Ownership-types-affine} to \prettyref{sec:Additional-thoughts},
the proposed new abstractions are described and examples using them
are given. Implementations of the examples are given in current OCaml
through a whole-program translation, that clarifies the computational
meaning of the abstractions and highlights the current limitations
of the runtime and type system. In \prettyref{sec:Comparison-with-existing},
the proposal is compared to existing PLs.

\subsection*{Acknowledgements}

Thanks to Raphaël Proust for introducing me to the topic of resource
management in PLs, and to Leo White for freely sharing thoughts on
the situation in OCaml. Thanks to Guillaume Combette: a milestone
for this proposal was an elementary reconstruction of RAII in LCBPV
which we obtained during his visit at LS2N, and which motivated several
aspects of it. Thanks to Frédéric Bour, Thomas Braibant, and François
Pottier for freely sharing their interest and experience with this
topic. Thanks to Gaëtan Gilbert, Adrien Guatto, Jacques-Henri Jourdan,
Gabriel Scherer, and Leo White, for comments and questions about an
earlier version of this proposal.

\section{\label{sec:Integrating-ownership-and}Integrating ownership and borrowing
with a GC}

Let us now describe the integration of a GC in an ownership-and-borrowing
system \emph{à la} C++\slash Cyclone\slash Rust following the definition
of the GC as a runtime optimisation for the collection of trivially
destructible values. (For the moment the unboxing optimisations are
not considered, but they do not fundamentally change the model, see
\prettyref{subsec:Unboxing}.)

\subsection{Owned, borrowed and GCed values}

There are GCed values typed with GC types, which never have a destructor.
They can be passed and returned freely. These are those already present
in OCaml.

Let us add types for resources, which are not managed by the GC but
with RAII, and which do have destructors that are called in a predictable
fashion. A value of the latter type is \emph{owned}. Owned values
can be moved, which transfers ownership, for instance to the caller,
to a callee, or to a data structure. This means that responsibility
for calling the destructor is transferred along with it. \emph{Ownership
types} combine in order to form other ownership types (for instance
a list of ownership type is an ownership type). Static analysis ensures
that a moved value is no longer accessed by the previous owner, for
instance with an affine type system.

In addition, owned values can be \emph{borrowed}. A borrowed value
is a copy which is given a \emph{borrow type}, denoting that the responsibility
of calling the destructor belongs to somebody else. Borrow types combine
to form other borrow types. There is no restriction on the amount
of times a borrowed value can be passed, but it should not be accessed
after the original value has been disposed of. This can be ensured
by a static analysis (typically inspired by type-and-effect systems
as introduced in \citealp{Tofte1994}); practical static analyses
combining this idea with ownership/linear types have already been
experimented in Cyclone and Rust (see in particular \citealp*{Fluet2006}).
Thus borrow types will need to carry annotations similar to Rust's
lifetimes.

There are now three modes of management:
\begin{description}
\item [{(G)}] \textbf{GCed values} and \textbf{GC types}
\item [{(O)}] \textbf{Owned values} and \textbf{ownership types}
\item [{(B)}] \textbf{Borrowed values} and \textbf{borrow types}
\end{description}
Each have pros and cons:
\begin{itemize}
\item GCed values can be copied freely, but cannot have destructors.
\item Owned values can only be moved, which allow them to support destructors,
to be used in a producer/consumer interaction (such as between components
of a program, or to receive and pass values from/to other runtimes),
to denote uniqueness, and to deal with large structures without impacting
the cost of tracing.
\item Borrowed values can be copied, but subject to the restriction that
it does not outlive the resource it originates from.
\end{itemize}
Such a diversity is not realistic without a plausible notion of resource
polymorphism:
\begin{itemize}
\item for the conciseness of the language,
\item for the expressiveness when mixing GC types and non-GC types,
\item at the level of types and their meaning, for simplicity and clarity
for the user, and
\item at runtime, for a simple and efficient implementation.
\end{itemize}
The example of C++11 shows that such a notion of resource polymorphism
also helps for backwards compatibility and extensibility of libraries.

The core part of the design is to understand GCed values polymorphically
both as borrowed values and as owned values.

\subsection{Resource polymorphism and runtime representation (level 3)}

RAII is a notion tightly integrated into the runtime. Let us start
there. In addition to traced pointers (with lowest bit set to 0),
let us use untraced pointers (with lowest bit set to 1). The latter
are allocated in the major heap and deallocated using RAII. An untraced
pointer can either be borrowed or owned. If borrowed, there is nothing
to deallocate. If owned, a compiler-generated destructor is called
at the end of the scope, which deallocates the memory.

The following invariant is maintained throughout: any live GCed value
is reachable either from the stack/registers, or from registered roots.
In order to use a GCed value as a sub-value of an owned value, the
GCed value is registered as a root at allocation. The compiler-generated
destructor is then in charge of unregistering the root. Thus RAII
is essential for the absence of leaks in the presence of exceptions.
If the sub-value is in the minor heap, the new pointer is registered
as a major-to-minor pointer and treated as such.

This leads to a first table for combining G, O, B types according
to whether the resulting value is allocated by GC or not (in order:
the strict pair, the type of lists, and the type of borrows, the latter
of which is an addition of the proposal):

\begin{table}[H]
\hspace*{\fill}
\begin{varwidth}{\linewidth}
\begin{tabular}{llllll}
\toprule 
\texttt{\emph{X}} & \texttt{\emph{X}}\texttt{{*}G} & \texttt{\emph{X}}\texttt{{*}O} & \texttt{\emph{X}}\texttt{{*}B} & \texttt{\emph{X}}\texttt{ list} & \texttt{\&}\texttt{\emph{X}}\tabularnewline
\midrule 
\texttt{G} & 0 & 1 & 0 & 0 & 0\tabularnewline
\texttt{O} & 1 & 1 & 1 & 1 & 1\tabularnewline
\texttt{B} & 0 & 1 & 0 & 0 & x\tabularnewline
\bottomrule
\end{tabular}\medskip{}

0: traced pointer (lowest bit set to 0)

1: untraced pointer (lowest bit set to 1)

x: either, same as original
\end{varwidth}
\hspace*{\fill}

\caption{\label{tab:Runtime-representation-in}Runtime representation for newly-created
values in function of the polarity}
\end{table}

In words:
\begin{itemize}
\item Structures comprised of owned values are allocated with RAII,
\item Discarding borrowed values is trivial, so structures comprised of
borrowed values and/or GCed values are allocated with the GC,
\item Any combination of GCed values and owned values is owned; RAII is
used to register and unregister the GCed value as a root as explained
above.
\end{itemize}
Thus, the runtime representation of structures alternates GCed phases
and non-GCed phases: a GCed value can contain a non-GCed value by
borrowing, and a non-GCed value can contain GCed values by rooting.
Notice that without borrowing, the heap would have a much simpler
structure consisting of a RAII phase with leaves pointing to a GCed
phase. From this angle it is clear that borrowing is essential for
expressiveness.

Passing (\emph{copying}) a borrowed or GCed value is done by copying
the pointer. Owned values cannot be copied; instead, passing (\emph{moving})
an owned value involves copying the pointer and setting the original
to zero. Recording the move by setting the pointer to zero is required
because destructor calls are determined statically, and since resources
are affine, it cannot be known statically whether a resource has moved:
for instance there can be branching code paths in which only one path
moves the resource. Destructors therefore need to know at runtime
whether a resource has moved. Of course, this imposes a linear (affine)
treatment of owned values.

Thus, for each type, the compiler generates a destructor (whose representation
can be defunctionalised), which: 1) tests for zero to detect whether
the variable has moved, 2) if not, applies user-supplied destructors,
and 3) deallocates. On the back-end, modular implicits (or at least
their back-end) can be re-used for generating the destructor, which
in turn allows the proposal to scale to abstract types and polymorphic
functions: functions polymorphic in an Ownership type variable take
an implicit module as an argument, which contains the compiler-generated
destructor.

Reference counting can be used for optimising the case where several
roots point to the same value. However, this is not the form of reference
counting that has been criticised for garbage collection. It avoids
the well-known drawbacks of reference counting: the difficulty to
collect cycles, the up-front cost, and cascades of reference count
updates. Indeed, the collection is still ultimately performed by tracing,
and the cascades are avoided for two reasons:
\begin{enumerate}
\item only the pointer at the interface between the RAII phase and the GCed
phase is reference-counted, and
\item copying roots can only happen by borrowing, and in this case it is
not necessary to update the reference count since the reference does
not outlive its resource.
\end{enumerate}
All in all, GCed values are determined to be reachable by a mix of
tracing and reference counting. There was a definite influence in
the design of this proposal from the thesis that all GCs lie in a
spectrum between tracing and reference counting \citep{Bacon2004}.

\subsection{Resource polymorphism, types, and meaning (level 2)}

Let us now provide an explanation to the runtime model in terms of
types.

A GCed value is typed by a GC type, an owned value by an ownership
type and a borrowed value by a borrow type. The mode of resource management
(or polarity) of a type is determined by induction according to the
following polarity table:

\begin{table}[H]
\hspace*{\fill}
\begin{varwidth}{\linewidth}
\begin{tabular}{llllll}
\toprule 
\texttt{\emph{X}} & \texttt{\emph{X}}\texttt{{*}G} & \texttt{\emph{X}}\texttt{{*}O} & \texttt{\emph{X}}\texttt{{*}B} & \texttt{\emph{X}}\texttt{ list} & \texttt{\&}\texttt{\emph{X}}\tabularnewline
\midrule 
\texttt{G} & \texttt{G} & \texttt{O} & \texttt{B} & \texttt{G} & \texttt{G}\tabularnewline
\texttt{O} & \texttt{O} & \texttt{O} & \texttt{O+B} & \texttt{O} & \texttt{B}\tabularnewline
\texttt{B} & \texttt{B} & \texttt{O+B} & \texttt{B} & \texttt{B} & \texttt{B}\tabularnewline
\bottomrule
\end{tabular}\medskip{}

where:
\begin{itemize}
\item \texttt{O} signifies linearity restrictions,
\item \texttt{B} signifies lifetime restrictions,
\item \texttt{G} signifies no restrictions.
\end{itemize}
\end{varwidth}
\hspace*{\fill}

\caption{Usage restrictions in function of the polarity}
\end{table}

In words:
\begin{itemize}
\item The modes G, O, B are closed under constructions.
\item A GC type can be used to form both ownership types (in combination
with ownership types) and borrow types (in combination with borrow
types).
\item The type of borrows is a borrow type, except for the type of borrows
to a GC type which is a GC type (itself, obviously).
\end{itemize}
In other words, a GC type can be seen both as an ownership type and
a borrow type. Indeed:
\begin{itemize}
\item \emph{A GCed value is owned} in the sense that holding the pointer
(i.e. copying it to the stack) is sufficient to prolong the life of
the value. Moreover, GCed values can be moved in the same way as owned
values are moved. The fact that GCed values can be used like owned
values can therefore be reflected in the language, so that it is possible
to give a single resource-polymorphic implementation to the function
which takes two lists as argument and merges them, for instance.
\item \emph{A GC type can also be seen as a borrow type}, in the sense that
GCed values can be copied without restriction. There is no difference
between a GCed value and a resource with trivial destructor allocated
at the largest region of the program, if the GC is considered as an
optimisation anticipating its collection.
\end{itemize}
For instance, a GCed value can be extracted from inside a borrowed
value and passed in an owned context, prolonging its lifetime.

Lastly, notice that if no borrowed or ownership types appear in a
type, then it is GC. Substituting ``GC'' with ``default-copiable'',
this is the same design that made the addition of non-copiable classes
in C++11 backwards-compatible with C++98.

\section{\label{subsec:Aside-Origins-of}Aside: history of move semantics
and resource polymorphism}

\subsection{The promises of linear logic}

It has been suggested from the beginning that the design and implementation
of functional programming languages could take inspiration from linear
logic \citep{Gir87} and its decomposition of intuitionistic logic.
\citet{Lafont1988} took inspiration from intuitionistic linear logic
to propose the mixing of strict and lazy evaluation as well as GC-less
automatic memory allocation, and justify in-place update of linear
values. Safety of parallel and concurrent programs by means of static
typing has been another promise of linear logic \citep{Abramsky1993}.
These applications are closely related to continuation-passing-style
models \citep{Berdine2000} and syntactic control of interference
\citep{Reynolds1978,OHearn1999}.

In a series of visionary articles, \citet{Baker1994,Baker1994a,Baker1995}
has proposed the integration in functional programming of concepts
and implementation techniques from systems programming, using abstractions
directly inspired from linear logic and supported by an extensive
bibliography of implementation techniques spanning more than three
decades. He described many ideas that are now at the basis of the
C++11 and Rust resource-management models.

Moving values, \citet{Baker1994} argues, is a more fundamental operation
than copying, and can be implemented by permutations of the stack
reminiscent of the structural rules of linear logic. Copying should
be explicit, or disabled when meaningless. It is noticed that this
linear treatment supports C++-like destructors, and helps avoid synchronisation.
It is also noticed that tail-call optimisation, far from being hindered,
is in fact the default behaviour in this model. 

\citet{Baker1994a} describes what is essentially the modern usage
of reference-counted pointers in C++11 and Rust, in which an alternation
of moving, borrowing, and deferred copying, is used to minimise reference-count
updates. Moreover it is suggested that similar linearity considerations
can also be useful for tracing GCs.

In \citet{Baker1995}, linear values are advocated as a modular abstraction
for resources in the sense of systems programming. Swapping with an
empty value is mentioned as an alternative to permuting the stack.
Linear abstract data types are proposed as a way to enforce the linearity
of types that mix linear and non-linear components by protecting the
underlying representation. Linearity of continuations justifies efficient
implementations of control operators. The compatibility of the model
with exceptions and non-local exits similarly to C++ destructors is
mentioned, as well as the added expressiveness of moving resources
compared to old C++. And more: only limited accounts of these rich
texts can be offered here.

It is clear, at least, that Baker has made the connection between
resource management and linear logic, including the compatibility
with RAII, and invented move semantics in the process.

Baker admits that the linear discipline is heavy. The only way to
pass an argument is to move it, and functions have to return unconsumed
arguments alongside the return value in a tuple. Borrowing, which
had been considered for reference counting, has not been considered
for resources. \citet{Minsky1996b} proposed the \emph{unique pointer},
precursor to the C++11 \emph{unique\_ptr}, which relaxes the move-only
discipline by allowing \emph{non-consumable parameters}, essentially
the possibility to pass the resource by copiable reference. In order
to ensure the absence of use after free, references to unique pointers
are subject to drastic syntactic usage restrictions. Static analyses
in the style of Cyclone \citep{Jim2002} had not emerged yet. \citet{Hinnant2002}
managed to integrate move semantics in a backwards-compatible extension
of C++ including \emph{unique\_ptr}.

These works of Baker were perhaps too in advance of their time. We
have found no mention of them in the rest of the literature about
applications of linearity in functional programming. Otherwise, when
they were mentioned, it was to succinctly point out their limitations.\footnote{\citealp{Minsky1996b,Clarke1998,Clarke2003}, and other articles remote
from the current discussion, mostly in the context of control of aliasing,
often in the context of object-oriented programming.} They do not appear to have been accounted for what they are: considerations
of language design for resource management, inspired by a connection
between foundational works in logic and the practice in PL and systems
implementations. Here, too, resource management is seen as more general
than control of aliasing, and linear logic inspires aspects of language
design that come before the type system.

\subsection{Resource polymorphism}

Many lessons on substructural type systems are summarised in \citet{Walker2005a},
such as the practical imperative of a notion of polymorphism for the
various linearity restrictions, or the interpretation of the exponential
modality of linear logic as reference-counting by \citet{Chirimar1996}.
The latter interpretation suggests an analogy between resource modalities
and \emph{smart pointers}, which hinted at a further understanding
of practical resource management from the point of view of Girard's
polarities.

To sum up, the proposed notion of resource polymorphism is supported
by three features:
\begin{enumerate}
\item Resource management modes are polarities.
\item Polarity tables define a polymorphism of data types.
\item A notion of subtyping between polarities extends polymorphism: especially,
GCed values are both owned and borrowed.
\end{enumerate}
This phrasing uses concepts from \citet{Gir91,Gir93}, but these three
features are at the basis of C++'s RAII and its later extension with
move semantics. For instance, (1.) In C++, default destructors and
copy operations are automatically defined, and (2.) in C++11 the non-copiable
character of a type is inherited. In addition, (3.) containers such
as \emph{std::vector} are polymorphic in the resource management mode:
copy operations are disabled at compilation when they are meaningless,
using the SFINAE idiom. These features are also at work in Rust (where
traits such as \emph{Copy}, \emph{Drop}, or \emph{Sized}, play the
role of polarities), and, beyond RAII-based languages, they can be
seen to some extent in Clean, ATS\footnote{\href{http://www.ats-lang.org}{http://www.ats-lang.org}, \citet{Zhu2005}.},
and others.

The view of GC types as simultaneously owning and borrowing can be
approximated with reference-counting pointers in C++ and Rust, although
this usage has the practical issues of reference-counted garbage collection,
in addition to being syntactically heavy.

Polarity tables can be considered an automatic and predictable selection
of the best resource-management mode for a value. In this sense, resource
polymorphism is a way to fill the \emph{static-automatic gap }\citep{Proust2016}
in the design space of resource management, using abstractions that
are compositional.

An originality of C++ and Rust's take on linearity is to emphasize
the \emph{how?} instead of the \emph{how many?}, by assigning a computational
contents to the copy, move, and drop operations. It is an old folklore
in linear logic that distinct exponential modalities can coexist,
so any interpretation in terms of \textquotedblleft counting the uses\textquotedblright{}
has to miss part of the message. Strikingly, RAII can be seen as arising
from shifting attention from \emph{can this value be disposed of?}
to \emph{how is this value to be disposed of?}.

In contrast, many investigations into linear type systems interpret
linearity as counting uses, starting with \citet{Wadler90lineartypes}.
Among the works that are of close interest to this proposal, this
is the case in \citet{Kobayashi1999}, \citet{Hofmann2000}, \citet{Shi2013}
and \citet{Tov2011}. Despite this limitation, they all present interesting
use cases of linear/affine types, such as capabilities, optimisations
and finer memory management. As an exception, there is a qualitative
(as opposed to quantitative) interpretation of substructural type
systems with type classes in \citet*{Gan2014}, which mentions the
analogy with C++'s custom copy and destruction operators, although
it misses developed examples, and is not designed for exception-safety.
In this proposal, the source of inspiration for paying attention to
the qualitative vs.\,quantitative aspects are constructions of actual
models of linear logic where this computational contents appears,
such as those of Bierman or Lafont (see \citealp{PAM2009}, for a
survey). From this angle, C++ move operators are analogous to monoidal
symmetry.

In C++ and Rust, parametric resource-polymorphism (3.) is obtained
with templates, with usual limitations and drawbacks (duplication,
non scalable to richer type systems, and in the case of C++, poor
error messages and slowness at compilation). In a language with proper
parametric polymorphism and abstract types, type variables must be
given polarities. Cyclone proposes a notion of subkinding \citep{Grossman2006},
and a similar feature closer to our context was further explored in
Alms \citep{Tov2011}. A notion of polymorphism of calling conventions,
expressed as a polymorphism of kinds (as opposed to types), was developed
in \citet{Eisenberg2017}, and later reused for \emph{multiplicity
polymorphism} in \citet*{Bernardy2018}.

The current proposal requires an approach that scales to the qualitative
interpretation of linearity; essentially, destructors need to be passed
when instantiating ownership type variables. Here, functions polymorphic
in an ownership type are proposed to depend on an implicit module
supplying the destructor. This idea, for which the design in \citeauthor{Grossman2006}'s
and \citeauthor{Tov2011}'s is the most fitting, was explored in \citet{Gan2014}
with type classes. In terms of a runtime model, this idea is also
similar in essence to the tag-free approach in \citet{Morrisett1995},
however applied to destructors only. \citet{Tov2011} is also a source
of inspiration for showing the existence of principal usage qualifiers
with the subkinding approach, and it will come back several times
in the rest of this proposal.

\section{\label{sec:Ownership-types-affine}Ownership types: affine types
with destructors}

Let us now describe additions to OCaml and give examples of uses.
The code given throughout is of two kinds.

\begin{lstlisting}[basicstyle=\sffamily]
(* Code aligned to the left: new syntax and examples
   for a language that does not exist yet. *)
\end{lstlisting}

\begin{lstlisting}[basicstyle=\sffamily\itshape,xleftmargin=8em]
(* Code indented to the right, in italics: a model of
   the proposed language, given by a whole-program
   translation in OCaml 4.06. *)
\end{lstlisting}
The model describes the observational behaviour, but does not respect
the runtime model: it is heavy and inefficient. Also it does not check
lifetimes, and linearity must be enforced by hand, so it does not
model a type system. It models level 2, the language abstractions.
It is similar in spirit to Stroustrup's dynamic model of ownership
alluded to in \citet{Stroustrup2015}: \emph{\textquotedblleft useless\textquotedblright },
\emph{\textquotedblleft inefficient\textquotedblright }, \emph{\textquotedblleft incompatible\textquotedblright },
but \emph{\textquotedblleft useful for thinking about ownership\textquotedblright }.
As such, it also underlines the limitations of the current OCaml runtime
and type system.

\subsection{Declaring a custom ownership type}

Consider a new type declaration, for affine types given as a pair
$(A,\delta)$ of a base type $A$ and a user-specified destructor
$\delta:A\rightarrow\text{unit}$.
\begin{lstlisting}[basicstyle=\sffamily]
type u = affine$(A,\delta)$
\end{lstlisting}
Let us make more precise what is meant with a pair $(A,\delta)$.
In Rust's terminology, consider a \emph{Drop }trait.
\begin{lstlisting}[basicstyle=\sffamily]
module type Droppable = sig
  type t
  val drop : t -> unit (* must not raise *)
end

type u = affine(M:Droppable)
\end{lstlisting}
Example: an input file.
\begin{lstlisting}[basicstyle=\sffamily]
module Droppable_in_channel : Droppable with type t = in_channel
  = struct
  type t = in_channel
  let drop = close_in_noerr
end

type file_in = affine(Droppable_in_channel)
\end{lstlisting}

This declaration creates a new type $\textsf{u}\mathrel{(=}\textsf{file\_in})$.
It has the same runtime representation as $\textsf{M.t}\mathrel{(=}\textsf{in\_channel})$.
However, being a new type has two consequences: one cannot pass an
\textsf{in\_channel} to a function that expects a \textsf{file\_in},
and one can define two different types of resource with different
destructors over the same base type.

We do the same in the dynamic model:
\begin{lstlisting}[basicstyle=\sffamily\itshape,xleftmargin=8em]
module type Droppable = sig
  type t
  val drop : t -> unit (* must not raise *)
end

module Droppable_in_channel : Droppable with type t = in_channel
  = struct
  type t = in_channel
  let drop = close_in_noerr
end
\end{lstlisting}
To each affine type, associate a module as follows:
\begin{lstlisting}[basicstyle=\sffamily\itshape,xleftmargin=8em]
module type Affine = sig
  type t
  (* ownership type *)
  type t_ref
  (* associated borrow type *)
  val create : t_ref -> t
  (* Create a resource *)
  val borrow : t -> t_ref
  (* Borrow a resource. Raises Use_after_move if the resource has been
     moved or destroyed meanwhile. If the borrowed value is used after
     the resource has been destroyed, this is an error which cannot be
     detected at runtime. In the proposed runtime that allocates memory
     with RAII, this can segfault, but in this model everything is GC'd
     so it will only violate user's invariants (we give an example
     later). *)
  val move : t -> t
  (* Move a resource. Using it on a non-live resource raises
     Use_after_free. Any copy not encapsulated in move can result in
     Double_free. *)
  val delete : t -> unit
  (* Compiler-generated destructor. Only to be called by RAII.scope. *)
end
\end{lstlisting}
The runtime of the dynamic model is as follows:
\begin{lstlisting}[basicstyle=\sffamily\itshape,xleftmargin=8em]
module RAII : sig
  type 'a ptr
  module Make (M : Droppable) : (Affine with type t_ref = M.t)
  (** Make a new ownership type from a droppable type *)
  val handle : (unit -> 'a) -> 'a
  (** Ensures that destructors are called in order when an exception
     escapes. The bodies of try..withs must be wrapped in a call to
     handle. Failure to do so results in leaks when raising an
     exception. *)
  val scope : (module Affine with type t = 'a) -> 'a -> ('a -> 'b) -> 'b
  (** Simulate a scope for a bound owned variable. Any resource not
     encapsulated in such a scope leaks. *)
end
  = struct
  type 'a ptr = 'a cell ref
  and 'a cell = Live of 'a | Moved | Freed
  type destructor_closure = (unit -> unit) Stack.t
  let dcs : destructor_closure Stack.t = Stack.create ()
  module S = Stack
  let push_closure () =
    S.push (S.create ()) dcs
  let destroy_closure () =
    S.iter (fun f -> f ()) (S.pop dcs)
  let push_destructor (f : unit -> unit) =
    S.push f (S.top dcs)
  let pop_destructor () : unit -> unit =
    S.pop (S.top dcs)
  let handle (f : unit -> 'a) : 'a =
    push_closure ();
    match f () with
    | x -> destroy_closure (); x
    | exception e -> destroy_closure (); raise e
  let _ = push_closure (); at_exit destroy_closure
  module Make (M : Droppable) = struct
    type t = M.t ptr
    type t_ref =  M.t
    let create x = ref (Live x)
    let borrow o = match !o with
      | Live x -> x
      | Moved ->
         failwith "use after move"
         (* This can happen when borrowing after a move. Avoided using:
            a linearity checker. *)
      | Freed ->
         failwith "use after free"
         (* This can happen after copying a resource. Avoided by: the
            language, by always moving. This does not account for all
            use-after-free bugs: the hard ones, which we cannot detect
            at runtime, are obtained when the borrow is created
            during the lifetime of the resource, but used after it is
            destroyed. The latter has to be avoided with a lifetime
            checker. *)
    let move o =
      let o' = create (borrow o) in o := Moved; o'
    let delete o = match !o with
      | Live x -> begin
          o := Freed;
          try M.drop x with
            _ -> ()
            (* An exception is raised while there could already stack
               unwinding for another raised exception. One can either
               call it undefined behaviour as in C++ and panic, or
               specify that all exceptions raised in destructors are
               silently ignored. *)
        end
      | Moved -> ()
      | Freed ->
         failwith "double free"
         (* This can happen after copying a resource. Avoided by: the
            language, by always moving. *)
  end
  let scope (type a) (module M : Affine with type t = a) x f =
    push_destructor (fun () -> M.delete x);
    let r = f x in
    pop_destructor () ();
    r
end
\end{lstlisting}
 An affine type is now declared as follows:
\begin{lstlisting}[basicstyle=\sffamily\itshape,xleftmargin=8em]
module File_in = RAII.Make (Droppable_in_channel)
type file_in = File_in.t
let scope_file_in f = RAII.scope (module File_in) f
\end{lstlisting}

\subsection{Creating an owned value}

To create a file\_in from an in\_channel, consider some new syntax:
\begin{lstlisting}[basicstyle=\sffamily]
new u(e : M.t)
\end{lstlisting}
Example: create an opened file and return it as a resource.
\begin{lstlisting}[basicstyle=\sffamily]
let open_file name : file_in = new file_in(open_in name)
\end{lstlisting}
Returning a resource transfers ownership of the resource to the caller.

\begin{lstlisting}[basicstyle=\sffamily\itshape,xleftmargin=8em]
let open_file name : file_in = File_in.create (open_in name)
\end{lstlisting}

An owned value is destroyed when it goes out of scope without being
moved.
\begin{lstlisting}[basicstyle=\sffamily]
let drop x = ()
(* all this function does is *not* using the resource x, which is
   already something: it destroys x. *)

let drop2 x y = ()
(* destroy in reverse order of creation (assuming right-to-left
   evaluation): x then y. *)

let fancy_drop x = try (let y = x in raise Exit) with Exit -> ()
(* x is moved into y, and y is dropped by the unwinding mechanism. *)
\end{lstlisting}

\begin{lstlisting}[basicstyle=\sffamily\itshape,xleftmargin=8em]
let drop x =
  scope_file_in x @@ fun x ->
  ()

let drop2' x y =
  scope_file_in x @@ fun x ->
  scope_file_in y @@ fun y ->
  ()
(* incorrect (for now), can you guess why? *)

let fancy_drop x =
  scope_file_in x @@ fun x ->
  try
    RAII.handle @@ fun () ->
    scope_file_in (File_in.move x) @@ fun y ->
    raise Exit
  with
  | Exit -> ()
\end{lstlisting}

\subsection{Moving an owned value}

An affine value can be moved but not copied.

\begin{lstlisting}[basicstyle=\sffamily]
let create_and_move name =
  let f = open_file name in
  drop f

let twice1 name =
  let f = open_file name in
  (f,f) (* typing error: f is affine *)

let twice2 name =
  let f = open_file name in
  drop f;
  f (* typing error: f is affine *)
\end{lstlisting}

\begin{lstlisting}[basicstyle=\sffamily\itshape,xleftmargin=8em]
let create_and_move name =
  scope_file_in (open_file name) @@ fun f ->
  drop (File_in.move f)

let twice1 name =
  scope_file_in (open_file name) @@ fun f ->
  (File_in.move f, File_in.move f)
(* failure "use after move" *)

let twice2 name =
  scope_file_in (open_file name) @@ fun f ->
  drop (File_in.move f);
  File_in.move f
(* failure "use after move" *)
\end{lstlisting}
If one copies instead of moving, one can have other kinds of errors.
The following two have no source equivalent:
\begin{lstlisting}[basicstyle=\sffamily\itshape,xleftmargin=8em]
let twice3 name =
  scope_file_in (open_file name) @@ fun f ->
  drop f; File_in.move f
(* failure "use after free". *)
let twice4 name =
  scope_file_in (open_file name) @@ fun f ->
  drop f; drop f
(* failure "double free". *)
\end{lstlisting}

\section{Borrowing}

A resource cannot be copied, but it can be borrowed. A borrowed value
can be copied without restriction, but it cannot be used after its
resource is destroyed. A borrowed value does not destroy the resource
when it goes out of scope. It is created with the following syntax:
\begin{lstlisting}[basicstyle=\sffamily]
&(x : t) : &t
\end{lstlisting}
The borrow type satisfies
\begin{lstlisting}[basicstyle=\sffamily]
&affine(M) = &M.t
\end{lstlisting}
and, for any GC type \textsf{t}:
\begin{lstlisting}[basicstyle=\sffamily]
&t = t
\end{lstlisting}
At runtime, \textsf{\&x} and \textsf{x} have the same representation.
When \textsf{x} is owned, the difference lies in \textsf{\&x} being
always copied and \textsf{x} being always moved.

\subsection{Example: safe reading from a file}

Example (compare with the equivalent one in \href{https://ocaml.org/learn/tutorials/file_manipulation.html}{https://ocaml.org/learn/tutorials/file\_manipulation.html}
which has one try/with and two explicit calls to \textsf{close\_in}):
\begin{lstlisting}[basicstyle=\sffamily]
let read_line name =
  let f = open_file name in
  (* if open_file raises an exception, no resource is created. *)
  print_endline (input_line &f);
  (* if input_line raises an exception, f is closed then. *)
  flush stdout
  (* f is closed then. *)
\end{lstlisting}

\begin{lstlisting}[basicstyle=\sffamily\itshape,xleftmargin=8em]
let read_line name =
  scope_file_in (open_file name) @@ fun f ->
  print_endline (input_line (File_in.borrow f));
  (* if input_line raises an exception, f is closed then *)
  flush stdout
  (* f is closed then *)
\end{lstlisting}

\subsection{Example: use-after-free}

Borrowing can induce more subtle bugs than linearity violations, which
requires to check that borrows do not outlive their resource. Cyclone
and Rust propose a compositional analysis inspired by type-and-effect
systems that assigns lifetime annotations to borrow types.

In OCaml, applying \textsf{input\_line} on a closed \textsf{in\_channel}
gives:
\begin{quote}
\emph{Sys\_error \textquotedbl Bad file descriptor\textquotedbl}
\end{quote}
In contrast, \textsf{open\_file} enforces that a \textsf{file\_in}
is always open (if it is not possible to open it, it safely raises
an exception before creating the resource). Thus, this error should
not arise.

First, we need to hide the definition of \textsf{file\_in}. Indeed,
as long as it is known that \textsf{\&file\_in = in\_channel}, then
it is possible to let the file handle escape in the old way, given
that \textsf{in\_channel} is GCed and can therefore be taken possession
of freely (it needs to be so, because we are interfacing with the
legacy OCaml library: this is just the unsafety of old \textsf{in\_channel}
surfacing).
\begin{lstlisting}[basicstyle=\sffamily]
module File : sig
  type file_in : O
  val open_file : string -> file_in
  val input_line : &file_in -> string
end = struct
  type file_in = affine(Droppable_in_channel)
  let open_file name = new file_in(open_in name)
  let input_line = Pervasives.input_line
end
\end{lstlisting}

The following program tries to use the resource after it has been
freed, and is expected to fail at compilation.
\begin{lstlisting}[basicstyle=\sffamily]
let use_after_free name =
  let f = File.open_file name in
  let g = &f in (* g has borrow type &File.file_in *)
  drop f;
  File.input_line g (* lifetime error: g outlives its resource. *)
\end{lstlisting}
The dynamic model does not perform this static analysis, and is therefore
allowed to violate the user's invariant:
\begin{lstlisting}[basicstyle=\sffamily\itshape,xleftmargin=8em]
let use_after_free name =
  scope_file_in (open_file name) @@ fun f ->
  let g = File_in.borrow f in
  drop (File_in.move f);
  input_line g (* Exception: Sys_error "Bad file descriptor". *)
\end{lstlisting}

\section{\label{sec:Polarity-tables-and}Polarity tables and pattern matching}

\subsection{Pair of owned}

Recall the polarity tables from Section 1. The pair \textsf{(x,y)}
is affine as soon as either \textsf{x} or \textsf{y} is affine. Its
compiler-generated destructor is obtained by combining those of \textsf{x}
and \textsf{y} in reverse order of creation (after testing for zero,
and before deallocating the cell).

Implicit modules are not available yet, so the dynamic model uses
plain modules with explicit instantiation.

\begin{lstlisting}[basicstyle=\sffamily\itshape,xleftmargin=8em]
module ATensor (P : Affine) (Q : Affine) : Affine
       with type t = P.t * Q.t
       with type t_ref = P.t_ref * Q.t_ref
  = struct
  type t = P.t * Q.t
  type t_ref = P.t_ref * Q.t_ref
  let create (x,y) = (P.create x, Q.create y)
  let borrow (x,y) = (P.borrow x, Q.borrow y)
  let move (x,y) = (P.move x, Q.move y)
  (* What about all the deep copies? In the actual proposed runtime
     model, moves and borrows only involve copying or moving the
     pointer. *)
  let delete (x,y) = P.delete x; Q.delete y
end
\end{lstlisting}

Example: in the following example, if the first \textsf{open\_file}
raises an exception, the second one is closed automatically. If the
result of this function is later dropped, both files are closed then.
\begin{lstlisting}[basicstyle=\sffamily]
let open2 name1 name2 = (open_file name1, open_file name2)
\end{lstlisting}
\begin{lstlisting}[basicstyle=\sffamily\itshape,xleftmargin=8em]
let open2 name1 name2 =
  let module File2 = ATensor (File_in) (File_in) in
  scope_file_in (open_file name2) @@ fun y ->
  scope_file_in (open_file name1) @@ fun x ->
  File2.move (x,y)
\end{lstlisting}

\subsection{Heterogeneous pair}

If the pair is heterogeneous, e.g. \textsf{x} is owned and \textsf{y}
is GCed or borrowed, then an implicit coercion of \textsf{y} from
GCed to owned is introduced, by registering a root and assigning a
destructor that unregisters the root. The runtime representation of
the value remains the same. These runtime details do not appear in
the model, and the coercion from GCed to affine here is trivial.
\begin{lstlisting}[basicstyle=\sffamily]
module Affine_of_GCd (M : sig type t end) : Affine
       with type t = M.t
       with type t_ref = M.t
  = struct
  type t = M.t
  type t_ref = M.t
  let create x = x
  let borrow x = x
  let move x = x
  let delete x = ()
end
\end{lstlisting}
One can pattern-match on an affine tensor with default destructor,
and this involves no additional operation compared to usual pattern-matching.
For instance, the following function takes an affine type and returns
the string:
\begin{lstlisting}[basicstyle=\sffamily]
let fst ((x,y) : string * file_in) : string = x
\end{lstlisting}
 Consistently with the type which indicates that the result is not
a resource, the result value can be copied without restrictions.

In the model, this has been expressed by defining \textsf{Tensor.t}
and \textsf{Affine\_of\_GCd.t} non-abstractly.
\begin{lstlisting}[basicstyle=\sffamily\itshape,xleftmargin=8em]
module Affine_of_String =
  Affine_of_GCd (struct type t = string end)
module String_and_file =
  ATensor (Affine_of_GCd (struct type t = string end)) (File_in)
let fst (z : string * file_in) : string =
  RAII.scope (module String_and_file) z @@ fun (x,y) ->
  x
\end{lstlisting}

When returning the second component instead, the resource \textsf{y}
is still alive after the destructor of \textsf{(x,y)} runs: indeed,
when that one runs, \textsf{y} has already moved, and in the proposed
runtime, all that the destructor sees in place of \textsf{y} is a
null pointer, which it ignores.
\begin{lstlisting}[basicstyle=\sffamily]
let snd ((x,y) : string * file_in) : file_in = y
\end{lstlisting}
\begin{lstlisting}[basicstyle=\sffamily\itshape,xleftmargin=8em]
let snd (z : string * file_in) : file_in =
  RAII.scope (module String_and_file) z @@ fun (x,y) ->
  File_in.move y
\end{lstlisting}

In contrast, one cannot pattern-match on a custom affine value whose
underlying type is a tensor, but only on borrows of such values.

\subsection{Sums}

One can define sum, option and list types similarly:
\begin{lstlisting}[basicstyle=\sffamily\itshape,xleftmargin=8em]
type ('a,'b) sum = Left of 'a | Right of 'b

module ASum (P : Affine) (Q : Affine)
       : Affine
       with type t = (P.t, Q.t) sum
       with type t_ref = (P.t_ref, Q.t_ref) sum
  = struct
  type t = (P.t, Q.t) sum
  type t_ref = (P.t_ref, Q.t_ref) sum
  let create = function
    | Left x -> Left (P.create x)
    | Right y -> Right (Q.create y)
  let borrow = function
    | Left x -> Left (P.borrow x)
    | Right y -> Right (Q.borrow y)
  let move = function
    | Left x -> Left (P.move x)
    | Right y -> Right (Q.move y)
  let delete = function
    | Left x -> P.delete x
    | Right y -> Q.delete y
end

(* more generally *)
module From_map (M : sig
             type 'a t
             val map : ('a -> 'b) -> 'a t -> 'b t
           end) (P : Affine) : Affine
       with type t = P.t M.t
       with type t_ref = P.t_ref M.t
  = struct
  type t = P.t M.t
  type t_ref = P.t_ref M.t
  let create x = M.map P.create x
  let borrow x = M.map P.borrow x
  let move x = M.map P.move x
  let delete x = let _ = M.map P.delete x in ()
end

module AOption = From_map (struct
                     type 'a t = 'a option
                     let map f = function
                       | Some x -> Some (f x)
                       | None -> None
                   end)

module AList = From_map (struct
                   type 'a t = 'a list
                   let map = List.map (* destroys in LIFO order *)
                 end)
\end{lstlisting}

\subsection{Example: a resource-safe interface to Mutex}

Using RAII one can ensure that all locks are released. This example
could be given earlier, except for \textsf{try\_unlock} which returns
an affine option type.

First we recall the mutex signature from the library.
\begin{lstlisting}[basicstyle=\sffamily\itshape,xleftmargin=8em]
module type Mutex_sig = sig
  type t
  val create : unit -> t
  val lock : t -> unit
  val try_lock : t -> bool
  val unlock : t -> unit
end
\end{lstlisting}
 RAII implementation:
\begin{lstlisting}[basicstyle=\sffamily]
module RAII_Mutex : sig
  (* GC'd type: a mutex is not a resource. Its life is prolonged by the locks
     holding it, so there is nothing to do on destruction. *)
  type t
  (* destroying a lock releases it *)
  type lock : O
  val create : unit -> t
  val lock : t -> lock
  val try_lock : t -> lock option (* affine *)
end
= struct
  type t = Mutex.t
  type lock = affine(struct
                       type t = Mutex.t
                       let drop = Mutex.unlock
                     end)
  let create = Mutex.create
  let lock m = Mutex.lock m; new lock(m)
  let try_lock m = if M.try_lock m then Some(new lock(m)) else None
end
\end{lstlisting}

\begin{lstlisting}[basicstyle=\sffamily\itshape,xleftmargin=8em]
module RAII_Mutex(Mutex : Mutex_sig) : sig
  type t
  type lock
  module Lock : Affine with type t = lock
  val create : unit -> t
  val lock : t -> lock
  val try_lock : t -> lock option
end = struct
  type t = Mutex.t
  module Lock = RAII.Make(struct
                    type t = Mutex.t
                    let drop = Mutex.unlock
                  end)
  type lock = Lock.t
  let create = Mutex.create
  let lock m = Mutex.lock m; RAII.create m
  let try_lock m = if Mutex.try_lock m then
                     Some(RAII.create m) (* scope is unnecessary
                                            because the context is
                                            pure *)
                   else None
end
\end{lstlisting}

\subsection{Example: try-locking a list of mutexes and releasing them reliably}

(While it can be polymorphic in the type of lockable values, it is
not done here for simplicity.)
\begin{lstlisting}[basicstyle=\sffamily]
module Locking : sig
  type t : O
  val try_lock : RAII_Mutex.t list -> t option
end
  = struct
  type t = RAII_Mutex.lock list (* by making t abstract, one ensures that the
                                   order does not change. *)
  let try_lock ms =
    let try_with_exn m = match RAII_Mutex.try_lock m with
      | Some l -> l
      | None -> raise Exit
    in
    try
      Some (List.rev_map try_with_exn ms)
    with
      Exit -> None
      (* all locked mutexes have been released *)
end
\end{lstlisting}
In particular the locks are released in reverse order. While not important
for mutexes, the ability to enforce the order of destruction can be
important for some other resources (think of transactions that need
to be rolled back).
\begin{lstlisting}[basicstyle=\sffamily\itshape,xleftmargin=8em]
module Locking (Mutex : Mutex_sig) : sig
  type t
  module T : Affine with type t = t
  val try_lock : RAII_Mutex(Mutex).t list -> T.t option
end
  = struct
  module M = RAII_Mutex (Mutex)
  module T = AList (M.Lock)
  module O = AOption (M.Lock)
  type t = M.lock list
  let scope x = RAII.scope (module T) x
  let try_lock ms =
    let try_with_exn (m : M.t) : M.lock =
      RAII.scope (module O) (M.try_lock m) @@ function
       | Some l -> M.Lock.move l
       | None -> raise Exit
    in
    try RAII.handle (fun () -> Some (List.rev_map try_with_exn ms))
    (* Wrong: assumes a ressource-aware List.rev_map, that releases
       partial lockings in case of failure. It is given further below. *)
    with Exit -> None
end
\end{lstlisting}

\subsection{Borrowed values and pattern-matching}

When borrowing a pair, one gets a pair of borrows. 
\begin{lstlisting}[basicstyle=\sffamily]
&(x : a * b) : &a * &b
&(x : string * file_in) : string * &file_in
\end{lstlisting}
In other words, \textsf{\&} is a homomorphism from affine to copiable
types. (Remember that \textsf{\&string = string} due to its G polarity.)
This allows us to pattern-match on borrows. 
\begin{lstlisting}[basicstyle=\sffamily]
match &(x : a * b) with (y : &a, z : &b) -> (z, y, z) : &b * &a * &b
match &(x : string * file_in) with (y : string, z : &file_in) -> (z, y) : &file_in * string
\end{lstlisting}

In the above example, the value of type \textsf{\&b {*} \&a {*} \&b},
of polarity B, is allocated with the GC (see \prettyref{tab:Runtime-representation-in}).
Thus, a pair of borrows can either be allocated with RAII, or with
the GC: the allocation method of a value of polarity B is not always
statically known. In particular,
\begin{lstlisting}[basicstyle=\sffamily]
string * &file_in $\neq$ string * in_channel
\end{lstlisting}
Indeed, \textsf{string {*} in\_channel} is of polarity G, and always
allocated with the GC, whereas \textsf{string {*} \&file\_in} can
be obtained by borrowing an owned pair. The equation 
\begin{lstlisting}[basicstyle=\sffamily]
&file_in = in_channel
\end{lstlisting}
 only holds in outermost position in the type, as in:
\begin{lstlisting}[basicstyle=\sffamily]
(fun (_ : string, z : &file_in) -> z) : string * &file_in -> in_channel
\end{lstlisting}
Other data types are treated similarly. The following are implicit
definitions in the language.
\begin{lstlisting}[basicstyle=\sffamily]
type ('a,'b) &sum = Left of &'a | Right of &'b (* = (&'a,&'b) sum *)
type 'a &option = Some of &'a | None (* = &'a option *)
type 'a &list = [] | (::) of &'a * 'a &list (* = &'a list *)
\end{lstlisting}

From the point of view of types, this design is likely to raise interesting
questions for type inference, with variants to investigate.

\subsection{Example: Zipper}

As an example of application of this design, it is possible to explore
an owned tree with a Zipper \citep{Huet1997a}. Its implementation
is not reproduced here because it is identical to the original one,
up to checking of lifetimes; we also assume it polymorphic in a sense
made more precise in \prettyref{sec:Parametric-resource-polymorphism}.
Then one can have an owned Zipper that takes ownership of the tree.
But one can also take a borrowed Zipper to explore the owned tree.
Then the initial Zipper is obtained at no cost by borrowing the owned
tree. It is therefore entirely allocated with RAII. Subsequent Zippers
are obtained by allocating new values with the GC, and therefore they
are allocated in part with the GC and in part with RAII. Thus, the
same polymorphic Zipper can be used both with owned and borrowed resources,
with in both cases the cost properties expected from a Zipper.

\subsection{Other data types}

Extending this approach to all OCaml data types raises interesting
questions, for instance with abstract types and GADTs. By adopting
a design à la \citet{Tov2011}, in which the polarity can depend on
type variables with an operator \textsf{<'a>} (\textquotedblleft the
polarity of \textsf{'a}\textquotedblright ), it is possible to declare
that an abstract type has the polarity of its argument: 
\begin{lstlisting}[basicstyle=\sffamily]
type 'a t : <'a>
\end{lstlisting}
or infer that the equality type is GCed even if it is an equality
between ownership types: 
\begin{lstlisting}[basicstyle=\sffamily]
type (_, _) eq (* : G *) = Refl : ('a : O, 'a : O) eq
\end{lstlisting}
It is also possible to reflect the polarity of phantom types in the
type of the GADT with the following idea:
\begin{lstlisting}[basicstyle=\sffamily]
type 'a gadt : <'a> =
  | GCd_value : ('b : G) -> unit gadt
  | Owned_value : ('b : O) -> owned_unit gadt
\end{lstlisting}
where \textsf{owned\_unit} is a dummy type of Owned polarity.
\begin{lstlisting}[basicstyle=\sffamily]
type owned_unit = affine(struct type t = unit  let drop () = () end)
\end{lstlisting}

It remains to be seen how polarities scale to all OCaml data-types;
but in case of stumbling blocks, polarity tables always leave the
option of disabling certain combinations.

\subsection{\label{subsec:Example-capabilities}Example: capabilities}

An example of application of an affine GADT is to encode an existential
type, and can be used to tie a capability to a data structure (although
this was not the original intent of the proposal). The following example
is from \citet{Tov2011}.
\begin{lstlisting}[basicstyle=\sffamily]
module CapArray : sig
  type ('a,'b) t
  type 'b cap : O
  type _ cap_array = Cap_array : ('a,'b) t * 'b cap -> 'a cap_array
  (* inferred as Ownership *)
  val make : int -> 'a -> 'a cap_array
  val set : ('a,'b) t -> int -> 'a -> 'b cap -> 'b cap
  val get : ('a,'b) t -> int -> 'b cap -> 'a * 'b cap
  val dirty_get : ('a,'b) t -> int -> 'a
end = struct
  type ('a,'b) t = 'a array
  type 'a cap = affine(struct type t = unit let drop _ = () end)
  type _ cap_array = Cap_array : ('a,'b) t * 'b cap -> 'a cap_array
  let make n x = Cap_array (Array.make n x, new cap())
  let set a n x _ = Array.set a n x
  let get a n _ = (Array.get a n, ())
  let dirty_get = Array.get
end
\end{lstlisting}
Nothing spectacular happens inside the dynamic model, because capabilities
do not require RAII.

\begin{lstlisting}[basicstyle=\sffamily\itshape,xleftmargin=8em]
module CapArray : sig
  type ('a,'b) t
  type 'b cap
  type 'b cap_ref (* One cannot do anything with a &cap, but we
                     need it for defining &cap_array *)
  module Cap (B : sig type b end) : Affine
         with type t = B.b cap
         with type t_ref = B.b cap_ref
  type _ cap_array = Cap_array : ('a,'b) t * 'b cap -> 'a cap_array
  type _ cap_array_ref =
    Cap_array_ref : ('a,'b) t * 'b cap_ref -> 'a cap_array_ref
  module Cap_Array (A : sig type a end) : Affine
         with type t = A.a cap_array
         with type t_ref = A.a cap_array_ref
  val make : int -> 'a -> 'a cap_array
  val set : ('a,'b) t -> int -> 'a -> 'b cap -> 'b cap
  val get : ('a,'b) t -> int -> 'b cap -> 'a * 'b cap
  val dirty_get : ('a,'b) t -> int -> 'a
end = struct
  type ('a,'b) t = 'a array
  type _ cap = unit
  type 'a cap_ref = 'a cap
  module Cap (B : sig type b end) = struct
    type t = unit
    type t_ref = unit
    let create () = ()
    let borrow () = ()
    let move () = ()
    let delete () = () (* a capability is affine and trivially destructible *)
  end
  type _ cap_array = Cap_array : ('a,'b) t * 'b cap -> 'a cap_array
  type _ cap_array_ref =
    Cap_array_ref : ('a,'b) t * 'b cap_ref -> 'a cap_array_ref
  let f : 'a cap -> (module Affine with type t = 'a cap) = fun (type a) x ->
    let module M = Cap (struct type b = a end) in
    (module M : Affine with type t = a cap)
  module Cap_Array (A : sig type a end) = struct
    type t = A.a cap_array
    type t_ref = A.a cap_array_ref
    let create = function Cap_array_ref (x,c) -> Cap_array (x,c)
    let borrow = function Cap_array (x,c) -> Cap_array_ref (x,c)
    let move x = x
    let delete x = ()
  end
  let make n (type a') (x : a') =
    let module M = Cap_Array (struct type a = a' end) in
    M.move (Cap_array (Array.make n x,()))
  let set ar n x () = Array.set ar n x
  let get ar n () = (Array.get ar n, ())
  let dirty_get ar n = Array.get ar n
end
\end{lstlisting}
The explicit threading in \textsf{set} and \textsf{get} is reminiscent
of the shortcoming of ownership without borrowing in \citet{Baker1995}.
A variation on this idea is to consider affine borrows \textsf{\&mut}
as in Rust. In Rust, \textsf{\&mut} is used for read-write operations
whereas \textsf{\&} is usually restricted to be read-only. Values
cannot be borrowed both with \textsf{\&mut} and \textsf{\&} at the
same time. Together with aliasing control provided by linearity, this
prevents data races on resources, and similar issues such as iterator
invalidation. There is no obstacle to including \textsf{\&mut} in
the current proposal. We come back to aliasing control in \prettyref{subsec:Languages-with-control}.

\section{\label{sec:Parametric-resource-polymorphism}Parametric resource
polymorphism}

Let us go back to the drop example. Make it polymorphic.
\begin{lstlisting}[basicstyle=\sffamily]
let drop (type a : O) (x : a) = ()
(* val drop : ('a : O).'a -> unit = <fun> *)
\end{lstlisting}
In the proposed runtime model, drop already knows how to move or borrow
resources of type \textsf{'a}, because these operations are the same
for all types. However, drop needs to know the compiler-generated
destructor for \textsf{'a}. Therefore, the universal quantification
on an affine type requires an implicit module supplying the destructor
(conversely, abstract types need to supply the destructor).

\begin{lstlisting}[basicstyle=\sffamily\itshape,xleftmargin=8em]
let drop (type a) (module A : Affine with type t = a) (x : a) =
  RAII.scope (module A) x @@ fun x ->
  ()
\end{lstlisting}
Since any GCed value can trivially be seen as an owned value, drop
above is also polymorphic in GCed values and behaves as expected.
In this case, a special null value for the destructor indicates at
runtime that there is no destructor to run and that allocations have
to be done with the GC.

For distinguishing between borrowed and owned variables during type
inference, one possibility is that by default inputs are inferred
to be borrows, and outputs to be owned. Given that GCed are polymorphically
borrowed and owned, this preserves the meaning of current polymorphic
code. Then the user can explicitly mark an owned input. Below, the
symbol \textsf{{*}} is used, but we avoid going into details of syntactic
choices. (See \emph{\textquotedblleft open questions\textquotedblright{}}
for more on this matter.)

\subsection{Example: merging two ordered lists}

\begin{lstlisting}[basicstyle=\sffamily]
let rec merge = function
    | list, []
    | [], list -> list
    | *h1::t1, *h2::t2 ->
        if &h1 <= &h2 then
          h1 :: merge (t1, h2::t2)
        else
          h2 :: merge (h1::t1, t2)
\end{lstlisting}
\begin{lstlisting}[basicstyle=\sffamily\itshape,xleftmargin=8em]
let merge (type a) (module A : Affine with type t = a) =
  let module AL = AList (A) in
  let module AL2 = ATensor (AL) (AL) in
  let rec iter x =
    RAII.scope (module AL2) x @@ function
     | list, [] | [], list -> AL.move list
     | h1::t1, h2::t2 ->
        if A.borrow h1 <= A.borrow h2 then
          A.move h1 :: iter (AL.move t1, AL.move (h2::t2))
        else
          A.move h2 :: iter (AL.move (h1::t1), AL.move t2)
  in
  iter
\end{lstlisting}

\subsection{Design and implementation}

The combination of polarities and parametric polymorphism gives rise
to interesting and crucial questions of type inference, principal
typing, and polarity specification for abstract types. These questions
are addressed by \citet{Tov2011} using subkinding and dependent kinds
(subtyping of polarities and polarities depending on the polarities
of type variables), extending an ML-like design. The hope for the
type system (level 1) is that it can be extended without major stumbling
blocks.

One question is whether all four polarities $\{G,O,B,O+B\}$ are needed
for type variables, or if only two polarities need to be presented
to the user (affine/copiable, as in Alms). The difference between
\textsf{'a:G} and \textsf{'a:B} is that the equation \textsf{\&t=t}
for \textsf{t:G} can be used during type-checking, e.g. \textsf{('a:G)
\&list = ('a:G) list}. It remains to be seen whether this is necessary
for expressiveness in concrete situations. But this question might
only be superficial: in all cases one wants that the lifetime of a
type \textsf{'a t : }$\left\langle \textsf{'a}\right\rangle $ is
deduced from the lifetime of \textsf{'a}.

As for the efficiency, one will likely want a guarantee that no efficiency
is lost due to passing a null destructor to polymorphic functions
every time a \textsf{G} is coerced to \textsf{O}. This case can be
optimised by treating polymorphic functions on a \emph{\textquotedblleft write
once, compile twice\textquotedblright{}} basis: whenever static knowledge
allows, one can call a specialised version that does not require the
implicit argument, that replaces moves with copies, and that always
allocates with the GC. The specialised version corresponds to what
OCaml would compile to currently. This optimisation is likely to apply
often: it will preserve the efficiency of current OCaml programs,
and will also apply in all cases involving borrow types.

\section{Owned mutable state}

Let us implement a resource-aware Stack module.

\subsection{Owned mutable cells}

Take OCaml's type of stacks:
\begin{lstlisting}[basicstyle=\sffamily]
type 'a t : <'a> = { mutable c : 'a list; mutable len : int; }
\end{lstlisting}
\textsf{'a} is now allowed to be a resource: polarity tables are the
same with or without the \textsf{mutable} keyword.

In addition, if \textsf{a} is owning, then the borrow type \textsf{a
\&t} is:
\begin{lstlisting}[basicstyle=\sffamily]
a &t = { &mutable c : a list; &mutable len : int; }
\end{lstlisting}
where \textsf{\&mutable} is a new keyword. In other words the computation
of \textsf{\&} stops at the \textsf{mutable} field. Both \textsf{mutable}
and \textsf{\&mutable} cells are lvalues of the specified type. In
addition, when used as an rvalue, the \textsf{\&mutable} cell has
the borrow type of its contents:
\begin{lstlisting}[basicstyle=\sffamily]
(x : 'a &t).c : 'a &list
\end{lstlisting}
such that \textsf{(\&x).c} and \textsf{\&(x.c)} are (intuitively)
equivalent.

As in Rust and Alms, the primitive operation is swapping between two
lvalues: 
\begin{lstlisting}[basicstyle=\sffamily]
s1.c <-> s2.c
\end{lstlisting}
Indeed, if the location contains a resource, swapping is more expressive
than \textsf{(<-)} because it does not involve any destruction of
resource. \textsf{(s1.c <- l)} can be expressed as follows: 
\begin{lstlisting}[basicstyle=\sffamily]
s1.c <-> (ref l).contents
\end{lstlisting}
A mutable cell owns its resource, and therefore \textsf{(<-)} proceeds
with the destruction of the previous value. (Unfortunately, defining
it as a function of \textsf{<->} is impossible without adding support
in OCaml for lvalues as arguments of functions).

\subsection{Example: Stack}

One idiom in Rust for dealing with mutable structures is to temporarily
take ownership of the contents, as with \textsf{take} and \textsf{swap}
below. Making use of this idiom in \textsf{push}, \textsf{pop}, and
\textsf{clear} is the only substantial difference with the original
OCaml \textsf{Stack} module.\footnote{One can also compare it with a similar data structure written in Rust:
\href{http://cglab.ca/~abeinges/blah/too-many-lists/book/first-final.html}{http://cglab.ca/$\sim$abeinges/blah/too-many-lists/book/first-final.html}.}

In the example below, \textsf{'a} is considered to be of most generic
polarity (O+B), although for backwards-compatibility one will need
to reserve \textsf{'a} for variables of polarity G, and find another
notation (\textsf{''a} or \textsf{'a : A}).

\begin{lstlisting}[basicstyle=\sffamily]
module Stack : sig
  type 'a t : <'a>
  val create : unit -> 'a t
  val swap : 'a &t -> 'a &t -> unit
  (** Exchange the contents of two stacks *)
  val take : 'a &t -> 'a t
  (** Take possession of the contents of the given stack.
      The argument is emptied in the process *)
  val push : 'a -> 'a &t -> unit
  val pop : 'a &t -> 'a
  val top : 'a &t -> &'a
  val clear : 'a &t -> unit
  val copy : 'a t -> 'a t
  val is_empty : 'a &t -> bool
  val length : 'a &t -> int
  val iter : (&'a -> unit) -> 'a &t -> unit
  val fold : ('b -> &'a -> 'b) -> 'b -> 'a &t -> 'b
end = struct
  type 'a t = { mutable c : 'a list; mutable len : int; }
  let create () = { c = []; len = 0; }
  let swap s1 s2 =
    s1.c <-> s2.c;
    s1.len <-> s2.len
  let take s =
    let s' = create () in
    swap s &s';
    s'
  let clear s = let _ = take s in ()
  let copy *s = { c = s.c; len = s.len; }
  let push *x s =
    let s' = take s in
    s.c <- x :: s'.c;
    s.len <- s'.len + 1
  let pop s =
    let s' = take s in
    match s'.c with
    | hd::tl -> s.c <- tl; s.len <- s'.len - 1; hd
    | []     -> raise Empty
  let top s =
    match s.c with
    | hd::_ -> hd
    | []     -> raise Empty
  let is_empty s = (s.c = [])
  let length s = s.len
  let iter f s = List.iter f s.c
  (* anticipating a resource-aware iter implementation *)
  let fold f acc s = List.fold_left f acc s.c
  (* anticipating a resource-aware fold_left implementation *)
end
\end{lstlisting}

In the dynamic model, let us implement a resource-polymorphic \textsf{ref}
following this model. It is essentially \textsf{aref} from Alms enriched
with RAII support and a \textsf{swap} function that operates on borrowed
values.
\begin{lstlisting}[basicstyle=\sffamily]
type 'a ref = { mutable contents: 'a }
let ref *x = { contents = x }
(* val ref : ('a : O).'a -> 'a ref = <fun> *)
let (:=) r x = r.contents <- x
(* val ( := ) : ('a : O).'a &ref -> 'a -> unit = <fun> *)
let (!) *r = r.contents
(* val ( ! ) : ('a : O).'a ref -> 'a = <fun> *)
let (!&) r = r.contents
(* val ( !& ) : ('a : O).'a &ref -> &'a = <fun> *)
let swap r r' = r.contents <-> r'.contents
(* val swap : ('a : O).'a &ref -> 'a &ref -> unit = <fun> *)
\end{lstlisting}
\begin{lstlisting}[basicstyle=\sffamily\itshape,xleftmargin=8em]
module Mutable (P : Affine) : sig
  module T : Affine
  val ref : P.t -> T.t
  val (:=) : T.t_ref -> P.t -> unit
  val (!) : T.t -> P.t
  val (!&) : T.t_ref -> P.t_ref
  val swap : T.t_ref -> T.t_ref -> unit
end = struct
  type t' = { mutable contents: P.t }
  module T = struct
    type t = t'
    type t_ref = t
    let create r = r
    let borrow r = r
    let move r = { contents = P.move r.contents }
    let delete r = P.delete r.contents
  end
  let ref x = { contents = P.move x }
  let (!) r = P.move r.contents
  let (!&) r = P.borrow r.contents
  let (:=) r x =
    RAII.scope (module P) (!r) @@ fun _ ->
    r.contents <- P.move x
  let swap r r' =
    let x = r.contents in
    r.contents <- r'.contents;
    r'.contents <- x
end
\end{lstlisting}

\subsection{Unsafe}

For a performance-critical library such as \textsf{Stack}, it is preferable
to have an \textsf{unsafe...end} block à la Rust, and write the more
efficient: 
\begin{lstlisting}[basicstyle=\sffamily]
let push *x s = unsafe s.c <- x :: s.c end; s.len <- s.len + 1
\end{lstlisting}
In an unsafe block, the checking of lifetimes and ownership is reverted
back to the user. The programmer can (and is encouraged to) reason
about the correctness of their code by showing it equivalent to the
previous one that uses \textsf{take}. In addition to being more efficient,
this code coincides on GCed values with the one in OCaml's \textsf{Stack}
implementation. \textsf{unsafe...end} can be considered without altering
the rest of the resource-management model.

\section{Closures}

Recall \textsf{drop2}: 
\begin{lstlisting}[basicstyle=\sffamily]
let drop2 x y = ()
\end{lstlisting}
When will the following partial application destroy \textsf{x}? 
\begin{lstlisting}[basicstyle=\sffamily]
let f = drop2 x in ...
\end{lstlisting}
when drop2 is applied to \textsf{x}, or when the scope of \textsf{f}
ends? One criterion is that Currification preserves the meaning. Then,
partial application must have the meaning of: 
\begin{lstlisting}[basicstyle=\sffamily]
let drop2 x = fun y -> let _ = x in ()
\end{lstlisting}
and therefore drop \textsf{x} only once \textsf{y} has been passed.

\subsection{Owning closures}

As in C++/Rust, closures can take possession of their variables. A
function such as the above where the captured variable is not used
is not necessarily a fabricated corner case: in RAII it is common
to use guards, that is values whose sole purpose is to exist for the
duration of a scope and not be used otherwise.

These languages provide a syntax to decide whether a variable has
to be moved, copied, or borrowed in a closure. This is always expressible
with lets (as illustrated above), so the problem of expressing which
affine variables are captured is reduced to a question of syntactic
sugar (into which this proposal does not go).

A consequence of the closure owning their resources is that function
values have a polarity determined by the closure (and neither by their
argument nor by their return type). This is predicted by the linear
call-by-push-value model. In semantic terms, a linear call-by-value
function has polarised type:
\[
\Downarrow(X\multimap Y)
\]
(readers of Levy might write its non-linear version $G(X\rightarrow FY)$,
the $F$ is implicit because it can be deduced from the context).

Here $\Downarrow$ is the type constructor of closures. In linear
call-by-push-value, there are several types of closures: linear ($\Downarrow$),
copiable ($!$). The current OCaml function type, copiable at will,
is:
\[
X\rightarrow Y=\text{!}(X\multimap Y)
\]
This leads to the introduction of an affine function type, as in Alms,
or as \textsf{FnOnce} in Rust. Affine means that the function has
to be used at most once (but it can use its argument as many times
as the argument's polarity allows).
\[
X\rightarrow_{1}Y=\text{\ensuremath{\Downarrow}}(X\multimap Y)
\]

Thus \textsf{drop2} can be given type: 
\begin{lstlisting}[basicstyle=\sffamily]
drop2 : ('a : O) ('b : O). 'a -> 'b ->$_1$ unit
\end{lstlisting}
denoting that the second closure holds a resource which will be consumed,
in this case the first argument.

Now what if an affine function value only uses its closure by borrowing?
In that case, although the closure is affine, it makes sense to call
the function several times. A third function space $\rightarrow_{\textsf{A}}$
is introduced, and is similar to $\rightarrow_{1}$ except that borrowed
values of type \textsf{\&(a $\rightarrow_{\textsf{A}}$ b)} can be
applied to values of type \textsf{a} as well. This is \textsf{Fn}
in Rust. (Let us hope this is enough function spaces!)
\begin{lstlisting}[basicstyle=\sffamily\itshape,xleftmargin=8em]
module AFun : sig
  type ('a,'b) fn
  type ('a,'b) fn_ref
  type ('a,'b) fn_once
  module Fn (M : sig type a type b end) : Affine
         with type t = (M.a, M.b) fn
         with type t_ref = (M.a, M.b) fn_ref
  module FnOnce (M : sig type a type b end) : Affine
         with type t = (M.a, M.b) fn_once
  val make : ('a -> 'b) -> ('a,'b) fn
  val make_once : ('a -> 'b) -> ('a,'b) fn_once
  val move_into_once : (module Affine with type t = 'c) ->
                       'c -> ('c -> ('a,'b) fn_once) -> ('a,'b) fn_once
  val move_into : (module Affine with type t = 'c
                                  and type t_ref = 'd) ->
                  'c -> ('d -> ('a,'b) fn) -> ('a,'b) fn
  val app_ref : ('a,'b) fn_ref -> 'a -> 'b
  val app_once : ('a,'b) fn_once -> 'a -> 'b
end = struct
  module type Closure = sig
    module A : Affine
    val x : A.t
  end
  let closure_move (module M : Closure) = (module struct
                                             module A = M.A
                                             let x = A.move M.x
                                           end : Closure)
  let closure_delete (module M : Closure) = M.(A.delete x)
  type ('a,'b) fn = ('a -> 'b) * (module Closure) list
  type ('a,'b) fn_ref = 'a -> 'b
  type ('a,'b) fn_once = ('a,'b) fn
  module Fn (M : sig type a type b end) : Affine
         with type t = (M.a, M.b) fn
         with type t_ref = (M.a, M.b) fn_ref
    = struct
    type t = (M.a, M.b) fn
    type t_ref = M.a -> M.b
    let borrow (f,l) = f
    let move (f,l) = (f, List.map closure_move l)
    let delete (f,l) = List.iter closure_delete l
  end
  module FnOnce (M : sig type a type b end) : Affine
         with type t = (M.a, M.b) fn_once
    = Fn (M)
  let make f = (f,[])
  let make_once f = (f,[])
  let move_into_once (type c) (module A : Affine with type t = c)
        res closure =
    let x = A.move res in
    let (f,l) = closure x in
    let x_mod = (module struct
                   module A = A
                   let x = x
                 end : Closure)
    in
    (f,x_mod::l)
  let move_into (type c) (type d)
        (module A : Affine with type t = c and type t_ref = d)
        res closure =
    let x = A.move res in
    let (f,l) = closure (A.borrow x) in
    let x_mod = (module struct
                   module A = A
                   let x = x
                 end : Closure)
    in
    (f,x_mod::l)
  let app_ref = (@@)
  let app_once (type a') (type b') fn x =
    let module F = Fn (struct type a = a' type b = b' end) in
    RAII.scope (module F) fn @@ fun (f,_) ->
    f x
end

let drop2 x =
  scope_file_in x @@ fun x ->
  AFun.move_into_once (module File_in) x (fun x ->
      AFun.make_once (fun y ->
          (scope_file_in y @@ fun y -> ())))
(* : File_in.t -> (File_in.t, unit) AFun.fn_once *)
\end{lstlisting}

\subsection{\label{subsec:Pure-functions}Functions with static closures}

What is the meaning of LCBPV's negative $\multimap$ then? Interpret
it as the type of functions not yet wrapped into (dynamic) closures,
as a distinction reminiscent of the one between function pointers
and closures seen in C++ and Rust.

The introduction of the closure can be statically delayed until a
value is actually required for the function, that is, when the function
is passed to another function, wrapped into a data structure, or fully
applied. (A full application is one which results in an expression
with positive type.) At either of these points, the set of free variables
used in the expression is known statically, and therefore introducing
the closure can be done then.

There are at least three reasons for introducing the negative $\multimap$
function space. If the function uses a resource, introducing the closure
forces to move the resource. Introducing it early can move the resource
earlier than necessary. For instance:
\begin{lstlisting}[basicstyle=\sffamily]
let f *x =
  let g y = &x in
  h &x (* is x still live or has it moved into the closure for g already? *)
  g
\end{lstlisting}
Delaying the introduction of the closure until the last line lets
us accept this program.

The second reason is that by typing differently returned functions
that do not need a closure yet, one reduces the number of intermediate
closures in call-by-value, statically and compositionally, unlike
the (non-compositional) nested redex optimisation \citep{Danvy2005}
or in the (dynamic) ZINC machine \citep{Leroy1990}.

This idea sheds light on the right-to-left evaluation order of arguments
seen in the ZINC. Indeed, it naturally leads to a right-to-left order,
because in the double application: 
\begin{lstlisting}[basicstyle=\sffamily]
f e1 e2
\end{lstlisting}
the type of \textsf{(f e1)} is negative and requires waiting for the
value of \textsf{e2}, and because left-to-right evaluation: 
\begin{lstlisting}[basicstyle=\sffamily]
let x1 = e1 in ... let xn = en in f x1 ... xn
\end{lstlisting}
is not macro-expressible given that its size is linear in the number
of arguments. (The idea of relating optimisations of closure introduction
in call-by-value to call-by-push-value is not new, a sketch of a relationship
with the nested redex optimisation and the ZINC machine is given in
\citealp{Spiwak2014}, and a different computational description for
the CBPV arrow is sketched.)

The third reason is an extension of the second one, in the presence
of affine closures. \citet{Tov2011} have noticed that currified functions
tend to accumulate annotations, often in a predictable manner, e.g.:
\[
\forall\alpha\beta.(\alpha\rightarrow\beta\rightarrow_{\left\langle \alpha\right\rangle }t\rightarrow_{\left\langle \alpha\right\rangle +\left\langle \beta\right\rangle }u)
\]
where $\rightarrow_{\left\langle \alpha\right\rangle }$ denotes the
type of closures with the same polarity as $\alpha$. This makes dependent
kinds necessary in their approach. This happens because a function
obtained by currying just adds the variable to the closure, as opposed
to a function that performs some computation before returning a closure.
By introducing the closure in a delayed manner, such a currified function
can be typed without annotations:
\[
\forall\alpha\beta.(\alpha\multimap\beta\multimap t\multimap u)
\]
Then, the closure is created, its variables moved, and its polarity
determined, either after full application, or when one tries to pass
or return the result of a partial application.

\section{Tail-call optimisation and control operators}

Tail-call optimisation (TCO) suffers from a bad interaction with destructors.
Given that a resource must be destroyed at the end of a scope, it
has to remain on the stack and prevents any TCO.

There are three separate issues:
\begin{enumerate}
\item This can lead to surprises, when writing polymorphic code or during
refactoring. Hence, the user should be allowed to specify that a call
is expected to be tail, and get an error if this is not the case.
\item This can prevent desired tail calls. The solution is to have a convenient
way to express that resources must be destroyed before the call, rather
than after.
\item The last condition being met, one must be able to actually implement
TCO.
\end{enumerate}
1) and 2) are largely a matter of syntactic choices and 1) is implemented
in OCaml with the \textsf{{[}@tailcall{]}} attribute. 

\subsection{Example: List.rev\_map}

Below is a solution for 2) and 3) involving an operator \textsf{tail\_call}
that destroys remaining resources before doing a tail call. In the
dynamic model it is implemented with a combination of an exception
and usual TCO.

\begin{lstlisting}[basicstyle=\sffamily]
let rev_map (type a : O) (type b : O) (f : a -> b) l =
  let rec rmap_f accu = function
    | [] -> accu
    | *a::*l -> tail_call rmap_f (f a :: accu) l
  in
  rmap_f [] l
\end{lstlisting}

\begin{lstlisting}[basicstyle=\sffamily\itshape,xleftmargin=8em]
let rev_map (type a) (module A : Affine with type t = a)
      (type b) (module B : Affine with type t = b)
      (f : a -> b) l =
  let module AL = AList(A) in
  let module BL = AList(B) in
  let exception Tail_rec of BL.t * AL.t in
  let rmap_f accu x =
    RAII.scope (module BL) accu @@ fun accu ->
    RAII.scope (module AL) x @@ function
     | [] -> BL.move(accu)
     | a::l -> raise (Tail_rec (f (A.move a) :: BL.move(accu), AL.move l))
  in
  let rec tail_rec x y =
    try
      RAII.handle (fun () -> rmap_f x y)
    with
      Tail_rec (x',y') -> (tail_rec [@tailcall]) x' y'
  in
  tail_rec [] l
\end{lstlisting}
\begin{lstlisting}[basicstyle=\sffamily]
let test_rev_map () =
  let x = open_file "/tmp/dummy1" in
  let y = open_file "/tmp/dummy2" in
  let z = open_file "/tmp/dummy3" in
  let rev_files = rev_map (fun *x -> x)
  let l = rev_files (rev_files [x;y;z]) in
  (* print the first lines in order and close the files *)
  rev_map (fun *x -> &x |> input_line |> print_endline) l
\end{lstlisting}
\begin{lstlisting}[basicstyle=\sffamily\itshape,xleftmargin=8em]
let test_rev_map () =
  let x = open_file "/tmp/dummy1" in
  let y = open_file "/tmp/dummy2" in
  let z = open_file "/tmp/dummy3" in
  let z' = File_in.borrow z in
  let rev_files = rev_map (module File_in) (module File_in)
                    (fun x -> File_in.move x)
  in
  let module L = AList(File_in) in
  let scope = RAII.scope (module L) in
  let _ =
      scope [File_in.move x; File_in.move y; File_in.move z] @@ fun l ->
      scope (rev_files (L.move l)) @@ fun l' ->
      scope (rev_files (L.move l')) @@ fun l'' ->
      (* val l'' : File_in.t list = [<abstr>; <abstr>; <abstr>] *)
      rev_map (module File_in) (module Affine_of_GCd(struct type t = unit end))
        (fun x -> scope_file_in x @@
         fun x -> (x |> File_in.borrow |> input_line |> print_endline))
        (L.move l'')
  in
  (* check that files are closed *)
  print_endline (input_line z')
  (* Exception: Sys_error "Bad file descriptor". (expected) *)
\end{lstlisting}

\subsection{Affine control}

More generally, such a \textsf{tail\_call} operator (and other operators
such as \textsf{return}) can be expressed with a control operator,
i.e. a variant of Landin's J operator \citep{Landin1965J}. \citet{Griffin90aformulae-as-types}
has shown that Scheme's call-with-current-continuation (call/cc) can
be typed with:
\[
((\alpha\textsf{ cont})\rightarrow\alpha)\rightarrow\alpha
\]
One can give \textsf{cont} a polarity $O+B$ to prevent it from being
called several times, passed inside its own arguments, and passed
inside the returned value. This ensures that it can be implemented
efficiently with stack unwinding, similarly to Scheme's call-with-escape-continuation
(call/ec).  Such a linear call/cc would provide an efficient statically-typed
error-handling mechanism, alternative to exceptions.

One can also give \textsf{cont} the polarity $O$ in the following:
\[
((\alpha\textsf{ cont})\rightarrow\bot)\rightarrow\alpha
\]
with the requirement that the control operator starts a new stack
(fiber). This linear variant of Felleisen's $\mathcal{C}$ operator
could be more of use in the context of concurrency.

Thus we conjecture that with appropriate typing, such operators can
be implemented at runtime like exceptions or effect handlers, but
that a difference appears with typing, as exemplified in the implementation
of \textsf{tail\_rec} above, which has to rely on a local exception
declaration specific to the type. Furthermore, in the presence of
negative function spaces (\ref{subsec:Pure-functions}), the type
\textsf{cont} can be refined to provide higher-order access to the
stack:
\[
(\alpha\multimap\beta)\textsf{ cont}=\alpha\otimes(\beta\textsf{ cont})
\]
\citep{Munch14Involutive}. This allows expressing stack inspection,
with possible applications to debugging and security \citep{Clements2004,Cle06}.

Once a type system with lifetimes and linearity is in place, investigating
and implementing such control operators is a low-hanging fruit.

\section{Example: a reference-counted cache}

Reference-counting pointers such as \emph{shared\_ptr/weak\_ptr} in
C++11 and Rc/Arc in Rust are useful for sharing resources compositionally,
such as between threads. The resource is freed as soon as the reference
count reaches zero. Unlike tracing GCs, they are not restricted to
contain non-resource values. For instance, one can easily implement
a cache of resources in C++11 as a vector of weak pointers. For non-resources,
such a cache can already be implemented in OCaml using GCed weak pointers.

In C++11 and Rust, reference-counting pointers are implemented using
custom copy operators that increase the reference count. Customizing
copy operations was not explored for several reasons. This requires
to enrich the proposal with new polarities, which makes the language
more complex, and is likely to have a performance impact (in contrast
with languages that implement polymorphism with templates). Also,
it is accepted that custom copy and move operations are more error-prone,
and are recommended to be seldom used, only for defining new resource
modalities when the ones from the libraries do not fit (this is C++11's
\emph{rule of zero}, an idiom part of coding guidelines and consolidated
in the C++14 standard). For now, omitting customizable copy and move
operators still allows explicit copy operations, which can be streamlined
with implicit modules, and does not preclude future extensions with
new polarities if necessary, which would again be backwards-compatible.

Still, elementary structures for dealing with resources, such as caches,
have to be implementable in ways fully integrated with the language.
For instance, assume that one is given bindings to Qt's QFileSystemWatcher\footnote{\href{https://doc.qt.io/qt-5/qfilesystemwatcher.html}{https://doc.qt.io/qt-5/qfilesystemwatcher.html}},
a platform-independent front-end to file monitoring systems such as
Linux's Inotify. 
\begin{lstlisting}[basicstyle=\sffamily]
module QFileSystemWatcher : sig
  val addPath : string -> bool
  val removePath : string -> bool
  (...)
\end{lstlisting}
In order to monitor a file, one needs to add its path, and register
a callback (using a mechanism not shown above). Monitoring a file
has a cost (there is a limit to their number on some platforms) so
one wants to end the monitoring as soon as it is no longer needed.
This interface is an example of implicit resource-unsafety: indeed,
if two parts of a program want to monitor the same file at overlapping
moments, then the first one to call \textsf{removePath} will stop
the second one from being subsequently notified.

A solution is to define \emph{file monitor handles}, which uniquely
handle the presence of a file in the monitoring list, and \emph{file
monitor guards}, responsible for handling a certain callback, where
each handle keeps track of the number of guards for this file. Therefore,
not only we need a cache of file monitor handles, but it has to be
flexible enough to allow the definition of custom guards, that are
also responsible for de-registering a callback.

As an example, we describe the implementation of a reference-counted
cache for resources that does not require customizing copying. Guards
are used for establishing the reference count. Let us assume that
OCaml's Map module has been extended to contain owned values.
\begin{lstlisting}[basicstyle=\sffamily]
module Rc_Cache (Ord : Map.OrderedType) : sig
  type key = Ord.t
  (** The type of cache keys. *)
  type ('a : O) t : O
  (** The type of caches. *)
  type ('a : O) cached : O+B
  (** The type of guards for accessing a cached resource. Holding a guard
      prolongs the life of the resource. The lifetime of the guard is
      limited by that of its cache. *)
  val create : unit -> 'a t
  (** Create an empty cache *)
  val get_or_add : key -> (key ->$_1$ 'a) -> 'a &t -> 'a cached
  (** [get_or_add k f c] retrieves the current cached resource with key [k],
      or associates [k] to [f k] if it does not exist yet, and returns
      a guard to the cached resource. The resource created by [f k] is
      destroyed as soon as there are no more guards pointing to it. *)
  val get : key -> 'a &t -> ('a cached) option
  (** [get k c] retrieves the current cached resource with key [k] if it
      exists. *)
  val deref : 'a &cached -> &'a
  (** Borrow the cached resource. Its lifetime is limited by that
      of its guard. *)
  val copy : 'a &cached -> 'a cached
  (** Get a new guard from an extant guard, increasing the reference count. *)
end = struct
  type key = Ord.t

  module M = Map.Make (Ord)

  type ('a : O) t = ('a * int ref) M.t ref

  let create () = ref (M.empty ())

  let take cache =
    let *cache' = create() in
    cache <-> &cache';
    !cache'

  let decr k cache =
    match M.find k (!& cache) with
    | (_, r) -> begin
        decr r;
        if !r = 0 then
          cache := take cache |> M.remove k
      end
    | exception Not_found -> assert false

  type 'a cached = affine(struct
                       type t = &'a * key * &'a t
                       let drop (_, k, cache) = decr k cache
                     end)

  let deref (x, _, _) = x

  let get k cache =
    match M.find k (!& cache) with
    | (x, r) -> begin
        incr r;
        Some (new cached(x, k, &cache))
      end
    | exception Not_found -> None

  let get_unsafe k cache =
    match get k cache with
    | Some *cached -> cached
    | None -> assert false

  let get_or_add k f cache =
    match get k cache with
    | Some *cached -> cached
    | None -> begin
        cache := take cache |> M.add k (f k, ref 0);
        get_unsafe k cache
      end

  let copy (_, k, cache) =
    get_unsafe k cache
end
\end{lstlisting}

Remarks:
\begin{itemize}
\item To come back to our example with \textsf{QFileSystemWatcher}, guards
be customised to perform additional duties such as registering and
unregistering callbacks, by composition of ownership types.
\item No effort has been made for thread-safety. However, thanks to \textsf{Map
}being a persistent data structure, the cache can be made thread-safe
by replacing the two \textsf{ref} types in the definition of \textsf{Rc\_Cache.t}
with atomic references as proposed in ocaml-multicore, and implement
synchronisation in \textsf{take}. Further efforts are of course necessary
for enabling thread-safe mutation of the contents.
\item If separating \textsf{\&} into a read-only \textsf{\&} and a linear
\textsf{\&mut}, the compiler would worry about a possible iterator
invalidation in \textsf{decr} and in \textsf{get\_or\_add}. One therefore
needs to have extra \textsf{unsafe...end} around the assignments of
\textsf{cache}. The soundness of this module has to be established
by external means, in the spirit of \citet{Benton2016a,Jung2018}.
This is expected, because it implements a resource-management policy
which is finer-grained than generic considerations on linearity and
lifetimes.
\end{itemize}

\section{\label{sec:Additional-thoughts}Additional considerations}

\subsection{Interfacing with foreign runtimes}

Interfacing with foreign runtimes is made easier with RAII, even those
that are GCed, as already demonstrated by Rust. Foreign functions
can receive owned values together with their destructors (for instance
responsible for unregistering GC roots). This is independent from
whether the foreign runtime possesses abstractions for ownership:
ownership comes with responsibility.

Conversely, OCaml functions can receive ownership of foreign values.
In that case, the destructor generated by the interface (for instance
deallocation) is automatically called using RAII.

One benefit of managing the allocation of owned values with RAII is
that it could become possible to structurally map types and values
when exchanging between the two languages, for the whole phase of
the structure which is RAII-allocated, provided the language on the
receiving end is rich enough to express the memory layout of the other
language.

It would also be interesting to see if by combining the RAII mechanisms
of two languages with exceptions that support it, one can achieve
interoperability for exceptions independently of their implementation.
(For instance for interoperating C++ and OCaml.)

\subsection{\label{subsec:Unboxing}Unboxing}

In OCaml, data structures of size 1 can be unboxed. This does interfere
with custom affine types: indeed, checking whether a resource has
moved involves checking if the pointer has been set to zero. Therefore,
custom affine types have to be taken on boxed types.

Unboxing is the subject of current discussions in OCaml; interest
has notably been shown for generalizing unboxing and making it predictable.
Seeing things from the angle of polarities suggests a type-directed
notion of unboxing fully compatible with the current proposal (in
the style of \citealp{Eisenberg2017}) by considering a polarity of
statically-known and fixed memory size (like Rust's \emph{Sized} trait).
But there are obstacles, such as the cost of passing the size at runtime
to polymorphic functions, and the necessity of occasionally having
explicit boxing modalities in the types (so as to avoid the cost of
repeatedly boxing and unboxing at function call boundaries).

Unless it is determined that the benefits outweigh the costs, the
proof-theoretic angle stops short of providing a solution to boxing-related
performance issues in OCaml. Still, it suggests that the proposal
might adapt to any sensible design for unboxing.

\subsection{Efficient implementation of exception-handling}

Let us describe in broad strokes how the implementation of exception-handling
can be extended conservatively to support RAII, as illustrated by
the dynamic model. With each trap one keeps a map associating registers
or stack addresses to their destructors, called above the destructor
closure. It is maintained by pushing and popping it according to the
lexical scoping of owned variables. Notably it does not need to be
updated when a resource is moved, because the destructor already knows
this at runtime by testing for null pointers. For this reason the
destructor closure can simply be implemented as a stack, that does
not need to be accessed during non-exceptional execution, only pushed
and popped. Stack unwinding when an exception is being raised, on
the other hand, now also involves calling the destructors in orderly
fashion, before handling the exception.

The runtime overhead compared to the current implementation is proportional
to the number of scopes for owned variables: in particular there is
no overhead for current resourceless code. This is close to optimal,
given that a destructor runs at the end of every such scope in any
case. The same idea can apply in multicore with fibers. In essence,
each fiber owns a set of resources, knows which ones it owns, how
to destroy them, and in which order. With effect handlers, the destructors
run when the captured continuation is dropped.

An alternative implementation model is inspired by permutation stacks
\citep{Baker1994}. In this model, a stack of affine values would
be maintained separately from the stack of copiable values. Moving
is done by permuting elements in the stack rather than copying+nullifying,
which avoids keeping nulls for values no longer there. This is more
elegant because it replaces the destructor closure, but the cost of
swapping must be assessed. Again, the costs are only paid by resource-intensive
users.

\subsection{Test for liveness}

A naive affine type system based on affine logic will reject programs
similar to the following one even if \textsf{b()} is constant:
\begin{lstlisting}[basicstyle=\sffamily]
if b() then f(x) (* moving x *);
if not b() then g(x) (* error: liveness of x is unknown *)
\end{lstlisting}
Since it is known at runtime whether a resource has moved, expressiveness
can greatly be increased by adding a test for liveness that converts
the run-time information into a static information.
\begin{lstlisting}[basicstyle=\sffamily]
if b() then f(x) (* moving x *);
if live(x) && not b() then g(x) (* moving x *)
\end{lstlisting}
Situations where the contents of \textsf{x} may have moved, rather
than \textsf{x} itself, still have to be prevented statically.

\subsection{Assessing the use for RAII allocation}

In an alternative representation, everything can be allocated with
the GC, and destructors do not manage allocation. However, this change
impacts two use cases: 1) Foreign pointers can still be received as
resources, and OCaml values can still be passed to a FFI by registering
the root; however this prevents any form of structural mapping between
OCaml structures and foreign structures; 2) The purpose of treating
very large data structures as a resource is to reduce the GC load
and perform real-time collection. Moreover, mutation avoids the write
barrier of the GC and is therefore more efficient.

RAII memory allocation (as opposed to RAII for non-memory resources)
is a low-hanging fruit once destructors are in place, and it is likely
useful for specific situations, but its practical impact will need
to be assessed concretely.

\section{\label{sec:Comparison-with-existing}Comparison with existing PLs
in a nutshell}

While some functional PLs have support for timely and exception-safe
release of resources (F\#\footnote{\href{http://fsharp.org/}{http://fsharp.org/}. This is not the \emph{destructors}
which are finalizers, but the IDisposable interface together with
the \emph{use} binder. The back-end appears to support moving (\emph{use}
will check for null before destruction), but this does not seem to
be reflected in the language.}, Scala\footnote{\href{http://www.scala-lang.org/}{http://www.scala-lang.org/}. This
is implemented as a library (ARM). Monadic sequencing seems to be
used to ensure the absence of uses after free.}) and there are some idioms for this in OCaml, these languages lack
move semantics and resource polymorphism, which gives resources their
first-class citizenship in C++11 and Rust. A natural question which
we do not explore due to lack of familiarity is whether the current
proposal could realistically be applied to them. Let us now focus
on comparisons with designs that are state-of-art in this respect.

\subsection{Systems PLs (Cyclone, C++11, Rust)}

This proposal extends the OCaml philosophy and runtime model. The
aim is to improve the resource management of a general-purpose PL,
rather than creating a new language dedicated to systems programming.
For this reason, compared to systems PLs, it can do with a coarser-grained
management of memory representation, and propose GCed allocation for
non-resources as the default. In contrast, the semantic perspective
suggested that the resource-management mechanisms in C++11 and Rust
(RAII and move semantics) could well belong to a high-level, general-purpose
programming language.

The language the closest to the proposal in this domain is Rust. Leaving
aside the differences in purpose, the novelty compared to Rust is
to propose a tight integration of RAII with a tracing GC, both at
the level of types (meaning), and at the level of the runtime.

Could the same design be used as a basis for integrating a GC in C++
or Rust? The proposed design is built around the assumption of a tracing
GC based on a uniform representation of values, with a run-time tag
distinguishing traced pointers from non-traced data. An alternative
approach in Rust, proposed by Goregaokar and Layzell with a prototype
GC design and implementation\footnote{\href{https://github.com/Manishearth/rust-gc/}{https://github.com/Manishearth/rust-gc/}.},
considers a \emph{Trace }trait that provides run-time information
about the memory layout for those types that are GCed, reminiscent
of the tag-free approach \citep{Morrisett1995}. It is similar in
several aspects: reference-counting is used at the interface between
the linear and the GCed values to track roots, the use of traits provides
some degree of resource polymorphism, and borrowing is proposed as
a mean to avoid incrementing reference counts. The syntactic footprint
of using Gc pointers is similar to that of reference-counted (Rc)
pointers. One way to see it is as an improvement of the latter for
those types with trivial destructors, fixing cascades of reference-count
updates, latency, and leaks due to cycles. It remains to be seen whether
this approach can be made efficient.

In contrast, our proposal integrates with a GC that has already been
designed for efficiency. The tagged representation is helpful in avoiding
reference count updates when assuming ownership of a GCed value (not
just borrowing), stack scanning reduces the need for reference-counted
roots, and pointers that move during minor collection can be handled
without any added complexity. Moreover, greater integration of the
GC within the language, notably a uniform representation between traced
and untraced values, affords a broader notion of polymorphism, such
as  between GC and borrow types with GC-allocated borrowed values,
and resource-polymorphic algorithms and data structures. This is where
we believe another potential sweet spot lies for OCaml, with no equivalent
among existing PLs.

\subsection{Functional PLs with generic linear types (Linear Lisp, Clean, Alms)}

Linear or affine types have for a long time been associated with the
idea of counting uses, even though foundational investigations on
linear logic have from the beginning let us hope of richer interpretations.
\citet{Baker1994,Baker1994a,Baker1995}, however, saw in them the
opportunity to improve on more than three decades of PL and systems
implementation techniques (\prettyref{subsec:Aside-Origins-of}).
Yet the resulting language Linear Lisp did not survive, and the lessons
remained misunderstood or ignored. ATS is another language that explores
linearity for memory allocation.

As long as their role was merely to decorate programs by counting
uses, the applications of linear types remained limited. This works
well for linear in-place update, which is explored by Clean and ATS
for purity and efficiency. But another prediction of linear logic
(not exclusive to it) is well-accepted but not always presented as
a form of linear typing: the importance of strict types. In fact,
in the linear call-by-push-value model, linear types are, by reminiscence
of focusing and polarisation in linear logic, a refinement of strict
types (i.e. linear types are positive), and strict types in the style
of ML are linear types that are copyable and discardable.

In the area of linear type systems, the closest language to suit the
requirements of this proposal is \citeauthor{Tov2011}'s Alms, which
to us appears state-of-art in many type-system aspects. Compared to
Alms, this proposal adopts a different perspective about affine types.
Specifically, this proposal essentially extends Alms by attaching
destructors to affine types. The semantic point of view suggests that
affine types with destructors are the natural guise of linear types
in the presence of control, instead of affine types being at odds
with linear types. From this perspective, further connections are
drawn between some of the language features and idioms of C++11, and
Girard's polarisation. Using these connections, interesting design
and applications of affine types for OCaml can be extracted from Stroustrup's
RAII and Hinnant et al.'s move semantics. Specifically, not only one
retains the uses of Alms, but user-provided and compiler-generated
destructors unlock many interesting uses beyond counting, such as
custom resources, low-level resources, and RAII allocation for performance
and interoperability. This also lets us consider a borrowing mechanism,
which in Alms would be vacuous (at least in their copiable form).

This tends to validate Baker's prediction that linear types open up
a new dimension for functional PLs. They address actual reported shortcomings
of OCaml, with perhaps many uses remaining to be discovered.

\subsubsection*{\label{subsec:Aside-Linear-Haskell}Aside: Linear Haskell}

Linear Haskell \citep{Bernardy2018} seems to be an exception to the
necessity of strictness among linear type systems, with applications
such as pure in-place update. Linear Haskell is \emph{\textquotedblleft designed
around linear types based on linear logic\textquotedblright }, a notion
that the authors oppose to linearity \emph{à la} Alms ($\mathsection$6.1:
\emph{\textquotedblleft this does not match linear logic (there is
no such thing as a linear proposition)\textquotedblright }), and to
linearity \emph{à la} Rust ($\mathsection$6.3: \emph{\textquotedblleft we
could have retrofitted uniqueness types to Haskell. But several points
guided our choice of designing Linear Haskell around linear logic
instead\textquotedblright }). It proposes a linear typing for arguments
of functions, rather than linear types per se, in a way reminiscent
of the negative (call-by-name) interpretation in linear polarised
models (e.g. \citealp{MelTab10RessourceModalities}). Linearity is
interpreted as counting uses. 

One of the main claims in Linear Haskell is a backwards compatibility
on the same level of ambition as the C++11 move semantics proposals.
However, exceptions in Haskell are not addressed nor mentioned, despite
the fact that linearity as opposed to affineness is important for
resource-manage\-ment protocols ($\mathsection$2.3, $\mathsection$5.2),
one of the two proposed applications. In addition, a prospective use
case is to handle values from foreign runtimes, including GCed ones
like in Rust ($\mathsection$7.3). How it would avoid being impacted
negatively by exceptions is not detailed.

In light of this proposal, it is natural to ask whether exception-safety
can be implemented with destructors. Unfortunately, the meaning of
destructors is unclear in the absence of strict types to attach them
to: in a lazy language, the notions of lifetimes and scopes are not
as clearly related. Laziness both prevents the timely release of owned
resources, and tends to let borrowed resources escape, unless special
measures are taken to impose sequencing.  Furthermore, Linear Haskell
proposes a notion of resource polymorphism (multiplicity polymorphism)
which describes the mixing of linear and unrestricted arguments to
functions for a single resource modality. This is a unique design,
for which it is unclear whether it can be adapted for non-idempotent
resource modalities, and multiple resource modalities with distinct
computational behaviour, as required for this proposal.

Another prospective application of linear typing in Haskell is to
address an issue with unpredictable performances ($\mathsection$7.1).
At this point it is natural to ask whether it is possible to further
take inspiration from linear logic and retrofit Haskell with strict
types: they are known to produce more predictable performances, and
affine types with destructors would fit naturally. In theory nothing
prevents applications of linear types in Haskell as described in this
proposal; in practice it likely requires a bigger evolution of the
language and libraries. According to \citet{Bolingbroke2009}: \emph{``We
would like to expose the ability to use `strict' types to the compiler
user }{[}...{]}\emph{ but allowing strictness annotations to appear
in arbitrary positions in types appears to require ad-hoc polymorphism,
and it is not obvious how to go about exposing the extra generality
in the source language in a systematic way\textquotedblright }. Starting
from a language that already features strict types is an essential
aspect making the current proposal practical and realistic.

\subsection{\label{subsec:Languages-with-control}Languages with control of aliasing
(Vault, Mezzo, Rust)}

The notion of linearity developed here can be combined with type abstraction
to provide a syntactic control of interference \citep{Reynolds1978,OHearn1999},
reminiscent of linear abstract data types \citep{Baker1995} and external
uniqueness \citep{Clarke2003}. In this scenario, the corresponding
borrow type cannot appear in the interface: there can be as many borrows
as one wants, as long as they cannot be used they cannot interfere!
This is exemplified with reader/writer locks with capabilities in
the style of Alms (\prettyref{subsec:Example-capabilities}).

The Rust language distinguishes linear borrows (\textsf{\&mut)} from
read-only borrows (\textsf{\&}), a design which can be used to lift
the previous restriction on borrows when controlling aliasing. This
design has been shown to be a practical and scaleable approach to
solving issues such as data races. The latter issue will become all
the more important with ocaml-multicore, which clarifies the semantics
of data races and provides atomic references \citep{Dolan2018}, but
has to exhort programmers to \emph{``still strive to avoid races''}
using locks and atomics\footnote{\href{https://github.com/ocamllabs/ocaml-multicore/wiki/Memory-model}{https://github.com/ocamllabs/ocaml-multicore/wiki/Memory-model}}.

First, there is no obstacle to applying the Rust approach for preventing
data races on owned values. As for shared mutable state, OCaml already
tends to reduce its need compared to C++11 and Rust, thanks to an
efficient support of persistent data structures with sharing. Lastly,
although racy by itself, keeping GCed shared local state in the language
leaves the door open to its encapsulation in safe abstractions in
libraries.  The design space remains to be explored: this has not
been the goal of this proposal so far. However, the combination of
the ingredients we already have at hand promises at least an efficient
model for data-race prevention in the style of Rust, without the need
for compromising backwards-compatibility of the language. (Libraries,
though, might require some adaptation.)

In \prettyref{subsec:The-relevance-of}, it has been suggested that
first extending OCaml with resource-friendliness in a backwards-compatible
manner can in particular provide a richer playground for tackling
the issue of data races in the future. A modest goal can therefore
be to organise such a playground such that further drastic evolutions
of the language will not be necessary.

If necessary, the hierarchy of polarities can be later be revised
so that one can ensure that abstract ownership types guarantee uniqueness.
This is currently only the case for concrete types; indeed, any polymorphic
function currently is instantiable with a GC type. This was a deliberate
choice and showed in a first time a simple and more modest design
inspired by polarisation. The alternative is to consider an affine
polarity $A$ and set $G<:A$ and $O<:A$ instead of $G<:O$. This
refinement would notably allow to consider unrestricted linear in-place
update by considering any variable of polarity $O$ an lvalue. For
instance, $O$-polymorphic list concatenation could be implemented
in constant time as originally suggested in \citet{Lafont1988}.

In the future we would like to have a closer look at the design and
uses of Vault \citep{Faehndrich2002} and Mezzo \citep{Pottier2013,Balabonski2016}.
There are synergies with Mezzo, which is also designed around ownership.
It explores the design space for a rich type system with fine-grained
control of permissions and aliasing using OCaml as a back-end, whereas
this proposal focuses on a backwards-compatible resource-management
model for OCaml itself. This proposal has less expressive types, even
assuming a distinction between linear and read-only borrows à la Rust,
but offers lower-level and more expressive abstractions.

\section{Summary}
\begin{flushright}
\emph{}%
\begin{minipage}[t]{0.6\columnwidth}%
\begin{flushright}
\emph{``It is time to beat turnstiles into technology. The elegance
and efficiency of linear types should allow functional languages to
finally go `main-stream'.''}
\par\end{flushright}
\begin{flushright}
--- \citet{Baker1995}
\par\end{flushright}%
\end{minipage}\emph{\medskip{}
}
\par\end{flushright}

This is a proposal for a novel resource-management model compatible
with the OCaml philosophy and implementation model, based on a new
understanding of RAII, move semantics, and resource polymorphism.
It achieves a synthesis of ownership-and-borrowing and linear functional
programming, notably by formulating GC allocation as a runtime optimisation
for types with trivial destructor.

Details of a language design and of a runtime model are given, requiring
little change compared to the current runtime:
\begin{itemize}
\item introducing untraced pointers allocated with RAII,
\item registering destructors with the exception mechanism, and
\item packing destructors with abstract types.
\end{itemize}
As for types, it proposes a natural polymorphism between ownership,
borrow, and GC types. It also raises interesting questions for a safe
and practical type system, for which the Cyclone, Rust, and Alms languages
provide ample prior work, that will need to be combined.

This proposed extension of OCaml integrates an ownership-and-borrowing
model with a tracing GC, with the aim to improve its:

\subsection*{Safety}
\begin{itemize}
\item New types for resources with destructors can be declared, the destructor
is guaranteed to be run predictably and reliably, including in the
presence of exceptions.
\item Facilities can be provided with default or built-in destructors, such
as structures of resources or foreign values.
\item Affine closures, notably, are important for local control and effect
handlers.
\item The Rust model for eliminating data races is likely to apply, and
the proposal forms a rich basis with which to further explore issues
of control of aliasing in the future.
\end{itemize}

\subsection*{Efficiency}
\begin{itemize}
\item Data structures can occasionally be allocated with RAII rather than
the GC, offering an allocation method which is timely and reduces
GC load, suitable for very large data structures and real-time collection,
and for which assignments bypass the write barrier of the GC.
\item The implementation for the current OCaml fragment, in particular the
GC and the exception handling, is essentially unchanged and therefore
remains as efficient as before.
\item There are further opportunities for linearity-and lifetime-directed
optimisations of garbage collection to investigate \citep{Baker1994a,Asati2014}.
\end{itemize}

\subsection*{Interoperability}
\begin{itemize}
\item Foreign values can be received as owned resources, and conversely
ownership of values can be given to foreign runtimes.
\item The design supports structural mapping between types.
\end{itemize}

\subsection*{Expressiveness}
\begin{itemize}
\item Resource polymorphism allows the definition of complex mixtures of
resource and non-resource types.
\item Resource polymorphism enables resource-generic structures and algorithms.
\item It is backwards-compatible with existing OCaml code: polarity tables
ensure that the default management for a non-resource data structure
is the most generic one, either GC or unboxed.
\item It enables efficient control operators.
\item The expressiveness of the proposal beyond all its intended uses above
remains to be explored.
\end{itemize}
\bibliographystyle{ACM-Reference-Format}
\phantomsection\addcontentsline{toc}{section}{\refname}\bibliography{../../references}


\begin{thebibliography}{68}

%%% the Web address.

\ifx \showCODEN    \undefined \def \showCODEN     #1{\unskip}     \fi
\ifx \showDOI      \undefined \def \showDOI       #1{#1}\fi
\ifx \showISBNx    \undefined \def \showISBNx     #1{\unskip}     \fi
\ifx \showISBNxiii \undefined \def \showISBNxiii  #1{\unskip}     \fi
\ifx \showISSN     \undefined \def \showISSN      #1{\unskip}     \fi
\ifx \showLCCN     \undefined \def \showLCCN      #1{\unskip}     \fi
\ifx \shownote     \undefined \def \shownote      #1{#1}          \fi
\ifx \showarticletitle \undefined \def \showarticletitle #1{#1}   \fi
\ifx \showURL      \undefined \def \showURL       {\relax}        \fi
% invisible to TeX
\providecommand\bibfield[2]{#2}
\providecommand\bibinfo[2]{#2}
\providecommand\natexlab[1]{#1}
\providecommand\showeprint[2][]{arXiv:#2}

\bibitem[\protect\citeauthoryear{Abramsky}{Abramsky}{1993}]%
        {Abramsky1993}
\bibfield{author}{\bibinfo{person}{Samson Abramsky}.}
  \bibinfo{year}{1993}\natexlab{}.
\newblock \showarticletitle{{C}omputational {I}nterpretations of {L}inear
  {L}ogic}.
\newblock \bibinfo{journal}{{\em Theor. Comput. Sci.\/}} \bibinfo{volume}{111},
  \bibinfo{number}{1{\&}2} (\bibinfo{year}{1993}), \bibinfo{pages}{3--57}.
\newblock
\showDOI{%
\url{https://doi.org/10.1016/0304-3975(93)90181-R}}


\bibitem[\protect\citeauthoryear{Anderson, Bergstrom, Goregaokar, Matthews,
  McAllister, Moffitt, and Sapin}{Anderson et~al\mbox{.}}{2016}]%
        {Anderson2016}
\bibfield{author}{\bibinfo{person}{Brian Anderson}, \bibinfo{person}{Lars
  Bergstrom}, \bibinfo{person}{Manish Goregaokar}, \bibinfo{person}{Josh
  Matthews}, \bibinfo{person}{Keegan McAllister}, \bibinfo{person}{Jack
  Moffitt}, {and} \bibinfo{person}{Simon Sapin}.}
  \bibinfo{year}{2016}\natexlab{}.
\newblock \showarticletitle{Engineering the servo web browser engine using
  Rust}. In \bibinfo{booktitle}{{\em ICSE '16}}.
\newblock
\showDOI{%
\url{https://doi.org/10.1145/2889160.2889229}}


\bibitem[\protect\citeauthoryear{Asati, Sanyal, Karkare, and Mycroft}{Asati
  et~al\mbox{.}}{2014}]%
        {Asati2014}
\bibfield{author}{\bibinfo{person}{Rahul Asati}, \bibinfo{person}{Amitabha
  Sanyal}, \bibinfo{person}{Amey Karkare}, {and} \bibinfo{person}{Alan
  Mycroft}.} \bibinfo{year}{2014}\natexlab{}.
\newblock \showarticletitle{Liveness-Based Garbage Collection}. In
  \bibinfo{booktitle}{{\em Compiler Construction - 23rd International
  Conference, {CC} 2014, Held as Part of the European Joint Conferences on
  Theory and Practice of Software, {ETAPS} 2014, Grenoble, France, April 5-13,
  2014. Proceedings}} {\em (\bibinfo{series}{Lecture Notes in Computer
  Science})}, \bibfield{editor}{\bibinfo{person}{Albert Cohen}} (Ed.),
  Vol.~\bibinfo{volume}{8409}. \bibinfo{publisher}{Springer},
  \bibinfo{pages}{85--106}.
\newblock
\showDOI{%
\url{https://doi.org/10.1007/978-3-642-54807-9_5}}


\bibitem[\protect\citeauthoryear{Bacon, Cheng, and Rajan}{Bacon
  et~al\mbox{.}}{2004}]%
        {Bacon2004}
\bibfield{author}{\bibinfo{person}{David~F. Bacon}, \bibinfo{person}{Perry
  Cheng}, {and} \bibinfo{person}{V.~T. Rajan}.}
  \bibinfo{year}{2004}\natexlab{}.
\newblock \showarticletitle{A unified theory of garbage collection}. In
  \bibinfo{booktitle}{{\em Proceedings of the 19th Annual {ACM} {SIGPLAN}
  Conference on Object-Oriented Programming, Systems, Languages, and
  Applications, {OOPSLA} 2004, October 24-28, 2004, Vancouver, BC, Canada}},
  \bibfield{editor}{\bibinfo{person}{John~M. Vlissides} {and}
  \bibinfo{person}{Douglas~C. Schmidt}} (Eds.). \bibinfo{publisher}{{ACM}},
  \bibinfo{pages}{50--68}.
\newblock
\showDOI{%
\url{https://doi.org/10.1145/1028976.1028982}}


\bibitem[\protect\citeauthoryear{Baker}{Baker}{1994a}]%
        {Baker1994}
\bibfield{author}{\bibinfo{person}{Henry~G. Baker}.}
  \bibinfo{year}{1994}\natexlab{a}.
\newblock \showarticletitle{Linear logic and permutation stacks - the Forth
  shall be first}.
\newblock \bibinfo{journal}{{\em {SIGARCH} Computer Architecture News\/}}
  \bibinfo{volume}{22}, \bibinfo{number}{1} (\bibinfo{year}{1994}),
  \bibinfo{pages}{34--43}.
\newblock
\showDOI{%
\url{https://doi.org/10.1145/181993.181999}}


\bibitem[\protect\citeauthoryear{Baker}{Baker}{1994b}]%
        {Baker1994a}
\bibfield{author}{\bibinfo{person}{Henry~G. Baker}.}
  \bibinfo{year}{1994}\natexlab{b}.
\newblock \showarticletitle{Minimum Reference Count Updating with Deferred and
  Anchored Pointers for Functional Data Structures}.
\newblock \bibinfo{journal}{{\em {SIGPLAN} Notices\/}} \bibinfo{volume}{29},
  \bibinfo{number}{9} (\bibinfo{year}{1994}), \bibinfo{pages}{38--43}.
\newblock
\showDOI{%
\url{https://doi.org/10.1145/185009.185016}}


\bibitem[\protect\citeauthoryear{Baker}{Baker}{1995}]%
        {Baker1995}
\bibfield{author}{\bibinfo{person}{Henry~G. Baker}.}
  \bibinfo{year}{1995}\natexlab{}.
\newblock \showarticletitle{"Use-Once" Variables and Linear Objects - Storage
  Management, Reflection and Multi-Threading}.
\newblock \bibinfo{journal}{{\em {SIGPLAN} Notices\/}} \bibinfo{volume}{30},
  \bibinfo{number}{1} (\bibinfo{year}{1995}), \bibinfo{pages}{45--52}.
\newblock
\showDOI{%
\url{https://doi.org/10.1145/199818.199860}}


\bibitem[\protect\citeauthoryear{Balabonski, Pottier, and Protzenko}{Balabonski
  et~al\mbox{.}}{2016}]%
        {Balabonski2016}
\bibfield{author}{\bibinfo{person}{Thibaut Balabonski},
  \bibinfo{person}{Fran{\c{c}}ois Pottier}, {and} \bibinfo{person}{Jonathan
  Protzenko}.} \bibinfo{year}{2016}\natexlab{}.
\newblock \showarticletitle{The design and formalization of Mezzo, a
  permission-based programming language}.
\newblock \bibinfo{journal}{{\em ACM Transactions on Programming Languages and
  Systems (TOPLAS)\/}} \bibinfo{volume}{38}, \bibinfo{number}{4}
  (\bibinfo{year}{2016}), \bibinfo{pages}{14}.
\newblock


\bibitem[\protect\citeauthoryear{Barendsen and Smetsers}{Barendsen and
  Smetsers}{1996}]%
        {Barendsen1996}
\bibfield{author}{\bibinfo{person}{Erik Barendsen} {and} \bibinfo{person}{Sjaak
  Smetsers}.} \bibinfo{year}{1996}\natexlab{}.
\newblock \showarticletitle{Uniqueness Typing for Functional Languages with
  Graph Rewriting Semantics}.
\newblock \bibinfo{journal}{{\em Mathematical Structures in Computer
  Science\/}} \bibinfo{volume}{6}, \bibinfo{number}{6} (\bibinfo{year}{1996}),
  \bibinfo{pages}{579--612}.
\newblock


\bibitem[\protect\citeauthoryear{Benton, Hofmann, and Nigam}{Benton
  et~al\mbox{.}}{2016}]%
        {Benton2016a}
\bibfield{author}{\bibinfo{person}{Nick Benton}, \bibinfo{person}{Martin
  Hofmann}, {and} \bibinfo{person}{Vivek Nigam}.}
  \bibinfo{year}{2016}\natexlab{}.
\newblock \showarticletitle{Effect-dependent transformations for concurrent
  programs}. In \bibinfo{booktitle}{{\em Proceedings of the 18th International
  Symposium on Principles and Practice of Declarative Programming, Edinburgh,
  United Kingdom, September 5-7, 2016}},
  \bibfield{editor}{\bibinfo{person}{James Cheney} {and}
  \bibinfo{person}{Germ{\'{a}}n Vidal}} (Eds.). \bibinfo{publisher}{{ACM}},
  \bibinfo{pages}{188--201}.
\newblock
\showDOI{%
\url{https://doi.org/10.1145/2967973.2968602}}


\bibitem[\protect\citeauthoryear{Berdine, O'Hearn, Reddy, and
  Thielecke}{Berdine et~al\mbox{.}}{2000}]%
        {Berdine2000}
\bibfield{author}{\bibinfo{person}{Josh Berdine}, \bibinfo{person}{Peter~W
  O'Hearn}, \bibinfo{person}{Uday~S Reddy}, {and} \bibinfo{person}{Hayo
  Thielecke}.} \bibinfo{year}{2000}\natexlab{}.
\newblock \showarticletitle{Linearly used continuations}. In
  \bibinfo{booktitle}{{\em Proceedings of the Third ACM SIGPLAN Workshop on
  Continuations (CW'01)}}. Citeseer, \bibinfo{pages}{47--54}.
\newblock


\bibitem[\protect\citeauthoryear{Bernardy, Boespflug, Newton, Peyton~Jones, and
  Spiwack}{Bernardy et~al\mbox{.}}{2018}]%
        {Bernardy2018}
\bibfield{author}{\bibinfo{person}{Jean{-}Philippe Bernardy},
  \bibinfo{person}{Mathieu Boespflug}, \bibinfo{person}{Ryan~R. Newton},
  \bibinfo{person}{Simon Peyton~Jones}, {and} \bibinfo{person}{Arnaud
  Spiwack}.} \bibinfo{year}{2018}\natexlab{}.
\newblock \showarticletitle{Linear Haskell: practical linearity in a
  higher-order polymorphic language}.
\newblock \bibinfo{journal}{{\em {PACMPL}\/}} \bibinfo{volume}{2},
  \bibinfo{number}{{POPL}} (\bibinfo{year}{2018}), \bibinfo{pages}{5:1--5:29}.
\newblock
\showDOI{%
\url{https://doi.org/10.1145/3158093}}


\bibitem[\protect\citeauthoryear{Bierman}{Bierman}{1995}]%
        {Bierman}
\bibfield{author}{\bibinfo{person}{Gavin Bierman}.}
  \bibinfo{year}{1995}\natexlab{}.
\newblock \showarticletitle{{W}hat is a categorical model of {I}ntuitionistic
  {L}inear {L}ogic?}. In \bibinfo{booktitle}{{\em Proc. TLCA}} {\em
  (\bibinfo{series}{Lecture Notes in Computer Science})},
  Vol.~\bibinfo{volume}{902}. \bibinfo{publisher}{Springer-Verlag},
  \bibinfo{pages}{78--93}.
\newblock


\bibitem[\protect\citeauthoryear{Bolingbroke and Peyton~Jones}{Bolingbroke and
  Peyton~Jones}{2009}]%
        {Bolingbroke2009}
\bibfield{author}{\bibinfo{person}{Maximilian~C. Bolingbroke} {and}
  \bibinfo{person}{Simon~L. Peyton~Jones}.} \bibinfo{year}{2009}\natexlab{}.
\newblock \showarticletitle{Types are calling conventions}. In
  \bibinfo{booktitle}{{\em Proceedings of the 2nd {ACM} {SIGPLAN} Symposium on
  Haskell, Haskell 2009, Edinburgh, Scotland, UK, 3 September 2009}},
  \bibfield{editor}{\bibinfo{person}{Stephanie Weirich}} (Ed.).
  \bibinfo{publisher}{{ACM}}, \bibinfo{pages}{1--12}.
\newblock
\showDOI{%
\url{https://doi.org/10.1145/1596638.1596640}}


\bibitem[\protect\citeauthoryear{Chirimar, Gunter, and Riecke}{Chirimar
  et~al\mbox{.}}{1996}]%
        {Chirimar1996}
\bibfield{author}{\bibinfo{person}{Jawahar Chirimar}, \bibinfo{person}{Carl~A.
  Gunter}, {and} \bibinfo{person}{Jon~G. Riecke}.}
  \bibinfo{year}{1996}\natexlab{}.
\newblock \showarticletitle{Reference Counting as a Computational
  Interpretation of Linear Logic}.
\newblock \bibinfo{journal}{{\em J. Funct. Program.\/}} \bibinfo{volume}{6},
  \bibinfo{number}{2} (\bibinfo{year}{1996}), \bibinfo{pages}{195--244}.
\newblock
\showDOI{%
\url{https://doi.org/10.1017/S0956796800001660}}


\bibitem[\protect\citeauthoryear{Clarke and Wrigstad}{Clarke and
  Wrigstad}{2003}]%
        {Clarke2003}
\bibfield{author}{\bibinfo{person}{Dave Clarke} {and} \bibinfo{person}{Tobias
  Wrigstad}.} \bibinfo{year}{2003}\natexlab{}.
\newblock \showarticletitle{External Uniqueness Is Unique Enough}. In
  \bibinfo{booktitle}{{\em {ECOOP} 2003 - Object-Oriented Programming, 17th
  European Conference, Darmstadt, Germany, July 21-25, 2003, Proceedings}} {\em
  (\bibinfo{series}{Lecture Notes in Computer Science})},
  \bibfield{editor}{\bibinfo{person}{Luca Cardelli}} (Ed.),
  Vol.~\bibinfo{volume}{2743}. \bibinfo{publisher}{Springer},
  \bibinfo{pages}{176--200}.
\newblock
\showDOI{%
\url{https://doi.org/10.1007/978-3-540-45070-2_9}}


\bibitem[\protect\citeauthoryear{Clarke, Potter, and Noble}{Clarke
  et~al\mbox{.}}{1998}]%
        {Clarke1998}
\bibfield{author}{\bibinfo{person}{David~G. Clarke}, \bibinfo{person}{John
  Potter}, {and} \bibinfo{person}{James Noble}.}
  \bibinfo{year}{1998}\natexlab{}.
\newblock \showarticletitle{Ownership Types for Flexible Alias Protection}. In
  \bibinfo{booktitle}{{\em Proceedings of the 1998 {ACM} {SIGPLAN} Conference
  on Object-Oriented Programming Systems, Languages {\&} Applications {(OOPSLA}
  '98), Vancouver, British Columbia, Canada, October 18-22, 1998.}},
  \bibfield{editor}{\bibinfo{person}{Bj{\o}rn~N. Freeman{-}Benson} {and}
  \bibinfo{person}{Craig Chambers}} (Eds.). \bibinfo{publisher}{{ACM}},
  \bibinfo{pages}{48--64}.
\newblock
\showDOI{%
\url{https://doi.org/10.1145/286936.286947}}


\bibitem[\protect\citeauthoryear{Clements}{Clements}{2006}]%
        {Cle06}
\bibfield{author}{\bibinfo{person}{John Clements}.}
  \bibinfo{year}{2006}\natexlab{}.
\newblock {\em \bibinfo{title}{{P}ortable and high-level access to the stack
  with {C}ontinuation {M}arks}}.
\newblock \bibinfo{thesistype}{Ph.D. Dissertation}.
  \bibinfo{school}{Northeastern University}.
\newblock


\bibitem[\protect\citeauthoryear{Clements and Felleisen}{Clements and
  Felleisen}{2004}]%
        {Clements2004}
\bibfield{author}{\bibinfo{person}{John Clements} {and}
  \bibinfo{person}{Matthias Felleisen}.} \bibinfo{year}{2004}\natexlab{}.
\newblock \showarticletitle{A tail-recursive machine with stack inspection}.
\newblock \bibinfo{journal}{{\em {ACM} Trans. Program. Lang. Syst.\/}}
  \bibinfo{volume}{26}, \bibinfo{number}{6} (\bibinfo{year}{2004}),
  \bibinfo{pages}{1029--1052}.
\newblock
\showDOI{%
\url{https://doi.org/10.1145/1034774.1034778}}


\bibitem[\protect\citeauthoryear{Curien, Fiore, and Munch-Maccagnoni}{Curien
  et~al\mbox{.}}{2016}]%
        {CFM2015}
\bibfield{author}{\bibinfo{person}{Pierre-Louis Curien},
  \bibinfo{person}{Marcelo Fiore}, {and} \bibinfo{person}{Guillaume
  Munch-Maccagnoni}.} \bibinfo{year}{2016}\natexlab{}.
\newblock \showarticletitle{{A} {T}heory of {E}ffects and {R}esources:
  {A}djunction {M}odels and {P}olarised {C}alculi}. In \bibinfo{booktitle}{{\em
  Proc. POPL}}.
\newblock
\showDOI{%
\url{https://doi.org/10.1145/2837614.2837652}}


\bibitem[\protect\citeauthoryear{Danos, Joinet, and Schellinx}{Danos
  et~al\mbox{.}}{1997}]%
        {danos95new}
\bibfield{author}{\bibinfo{person}{Vincent Danos},
  \bibinfo{person}{Jean-Baptiste Joinet}, {and} \bibinfo{person}{Harold
  Schellinx}.} \bibinfo{year}{1997}\natexlab{}.
\newblock \showarticletitle{{A} {N}ew {D}econstructive {L}ogic: {L}inear
  {L}ogic}.
\newblock \bibinfo{journal}{{\em Journal of Symbolic Logic\/}}
  \bibinfo{volume}{62 (3)} (\bibinfo{year}{1997}), \bibinfo{pages}{755--807}.
\newblock


\bibitem[\protect\citeauthoryear{Danvy and Nielsen}{Danvy and Nielsen}{2005}]%
        {Danvy2005}
\bibfield{author}{\bibinfo{person}{Olivier Danvy} {and}
  \bibinfo{person}{Lasse~R. Nielsen}.} \bibinfo{year}{2005}\natexlab{}.
\newblock \showarticletitle{{CPS} transformation of beta-redexes}.
\newblock \bibinfo{journal}{{\em Inf. Process. Lett.\/}} \bibinfo{volume}{94},
  \bibinfo{number}{5} (\bibinfo{year}{2005}), \bibinfo{pages}{217--224}.
\newblock


\bibitem[\protect\citeauthoryear{Dolan, Sivaramakrishnan, and
  Madhavapeddy}{Dolan et~al\mbox{.}}{2018}]%
        {Dolan2018}
\bibfield{author}{\bibinfo{person}{Stephen Dolan}, \bibinfo{person}{KC
  Sivaramakrishnan}, {and} \bibinfo{person}{Anil Madhavapeddy}.}
  \bibinfo{year}{2018}\natexlab{}.
\newblock \showarticletitle{Bounding Data Races in Space and Time}. In
  \bibinfo{booktitle}{{\em Proc. PLDI 2018}}.
\newblock


\bibitem[\protect\citeauthoryear{Eisenberg and Peyton~Jones}{Eisenberg and
  Peyton~Jones}{2017}]%
        {Eisenberg2017}
\bibfield{author}{\bibinfo{person}{Richard~A. Eisenberg} {and}
  \bibinfo{person}{Simon Peyton~Jones}.} \bibinfo{year}{2017}\natexlab{}.
\newblock \showarticletitle{Levity polymorphism}. In \bibinfo{booktitle}{{\em
  Proceedings of the 38th {ACM} {SIGPLAN} Conference on Programming Language
  Design and Implementation, {PLDI} 2017, Barcelona, Spain, June 18-23, 2017}},
  \bibfield{editor}{\bibinfo{person}{Albert Cohen} {and}
  \bibinfo{person}{Martin~T. Vechev}} (Eds.). \bibinfo{publisher}{{ACM}},
  \bibinfo{pages}{525--539}.
\newblock
\showDOI{%
\url{https://doi.org/10.1145/3062341.3062357}}


\bibitem[\protect\citeauthoryear{F{\"{a}}hndrich and DeLine}{F{\"{a}}hndrich
  and DeLine}{2002}]%
        {Faehndrich2002}
\bibfield{author}{\bibinfo{person}{Manuel F{\"{a}}hndrich} {and}
  \bibinfo{person}{Robert DeLine}.} \bibinfo{year}{2002}\natexlab{}.
\newblock \showarticletitle{Adoption and Focus: Practical Linear Types for
  Imperative Programming}. In \bibinfo{booktitle}{{\em Proceedings of the 2002
  {ACM} {SIGPLAN} Conference on Programming Language Design and Implementation
  (PLDI), Berlin, Germany, June 17-19, 2002}},
  \bibfield{editor}{\bibinfo{person}{Jens Knoop} {and}
  \bibinfo{person}{Laurie~J. Hendren}} (Eds.). \bibinfo{publisher}{{ACM}},
  \bibinfo{pages}{13--24}.
\newblock
\showDOI{%
\url{https://doi.org/10.1145/512529.512532}}


\bibitem[\protect\citeauthoryear{Fluet, Morrisett, and Ahmed}{Fluet
  et~al\mbox{.}}{2006}]%
        {Fluet2006}
\bibfield{author}{\bibinfo{person}{Matthew Fluet}, \bibinfo{person}{Greg
  Morrisett}, {and} \bibinfo{person}{Amal~J. Ahmed}.}
  \bibinfo{year}{2006}\natexlab{}.
\newblock \showarticletitle{Linear Regions Are All You Need}. In
  \bibinfo{booktitle}{{\em Programming Languages and Systems, 15th European
  Symposium on Programming, {ESOP} 2006, Held as Part of the Joint European
  Conferences on Theory and Practice of Software, {ETAPS} 2006, Vienna,
  Austria, March 27-28, 2006, Proceedings}} {\em (\bibinfo{series}{Lecture
  Notes in Computer Science})}, \bibfield{editor}{\bibinfo{person}{Peter
  Sestoft}} (Ed.), Vol.~\bibinfo{volume}{3924}. \bibinfo{publisher}{Springer},
  \bibinfo{pages}{7--21}.
\newblock
\showDOI{%
\url{https://doi.org/10.1007/11693024_2}}


\bibitem[\protect\citeauthoryear{Gan, Tov, and Morrisett}{Gan
  et~al\mbox{.}}{2014}]%
        {Gan2014}
\bibfield{author}{\bibinfo{person}{Edward Gan}, \bibinfo{person}{Jesse~A. Tov},
  {and} \bibinfo{person}{Greg Morrisett}.} \bibinfo{year}{2014}\natexlab{}.
\newblock \showarticletitle{Type Classes for Lightweight Substructural Types}.
  In \bibinfo{booktitle}{{\em Proceedings Third International Workshop on
  Linearity, {LINEARITY} 2014, Vienna, Austria, 13th July, 2014.}} {\em
  (\bibinfo{series}{{EPTCS}})}, \bibfield{editor}{\bibinfo{person}{Sandra
  Alves} {and} \bibinfo{person}{Iliano Cervesato}} (Eds.),
  Vol.~\bibinfo{volume}{176}. \bibinfo{pages}{34--48}.
\newblock
\showDOI{%
\url{https://doi.org/10.4204/EPTCS.176.4}}


\bibitem[\protect\citeauthoryear{Girard}{Girard}{1987}]%
        {Gir87}
\bibfield{author}{\bibinfo{person}{Jean-Yves Girard}.}
  \bibinfo{year}{1987}\natexlab{}.
\newblock \showarticletitle{{L}inear {L}ogic}.
\newblock \bibinfo{journal}{{\em Theoretical Computer Science\/}}
  \bibinfo{volume}{50} (\bibinfo{year}{1987}), \bibinfo{pages}{1--102}.
\newblock


\bibitem[\protect\citeauthoryear{Girard}{Girard}{1991}]%
        {Gir91}
\bibfield{author}{\bibinfo{person}{Jean-Yves Girard}.}
  \bibinfo{year}{1991}\natexlab{}.
\newblock \showarticletitle{{A} new constructive logic: {C}lassical logic}.
\newblock \bibinfo{journal}{{\em Math. Struct. Comp. Sci.\/}}
  \bibinfo{volume}{1}, \bibinfo{number}{3} (\bibinfo{year}{1991}),
  \bibinfo{pages}{255--296}.
\newblock


\bibitem[\protect\citeauthoryear{Girard}{Girard}{1993}]%
        {Gir93}
\bibfield{author}{\bibinfo{person}{Jean-Yves Girard}.}
  \bibinfo{year}{1993}\natexlab{}.
\newblock \showarticletitle{{O}n the {U}nity of {L}ogic}.
\newblock \bibinfo{journal}{{\em Ann. Pure Appl. Logic\/}}
  \bibinfo{volume}{59}, \bibinfo{number}{3} (\bibinfo{year}{1993}),
  \bibinfo{pages}{201--217}.
\newblock


\bibitem[\protect\citeauthoryear{Griffin}{Griffin}{1990}]%
        {Griffin90aformulae-as-types}
\bibfield{author}{\bibinfo{person}{Timothy~G. Griffin}.}
  \bibinfo{year}{1990}\natexlab{}.
\newblock \showarticletitle{{A} {F}ormulae-as-{T}ypes {N}otion of {C}ontrol}.
  In \bibinfo{booktitle}{{\em Seventeenth Annual ACM Symposium on Principles of
  Programming Languages}}. \bibinfo{publisher}{ACM Press},
  \bibinfo{pages}{47--58}.
\newblock


\bibitem[\protect\citeauthoryear{Grossman}{Grossman}{2006}]%
        {Grossman2006}
\bibfield{author}{\bibinfo{person}{Dan Grossman}.}
  \bibinfo{year}{2006}\natexlab{}.
\newblock \showarticletitle{Quantified types in an imperative language}.
\newblock \bibinfo{journal}{{\em {ACM} Trans. Program. Lang. Syst.\/}}
  \bibinfo{volume}{28}, \bibinfo{number}{3} (\bibinfo{year}{2006}),
  \bibinfo{pages}{429--475}.
\newblock
\showDOI{%
\url{https://doi.org/10.1145/1133651.1133653}}


\bibitem[\protect\citeauthoryear{Grossman, Morrisett, Jim, Hicks, Wang, and
  Cheney}{Grossman et~al\mbox{.}}{2002}]%
        {Grossman2002}
\bibfield{author}{\bibinfo{person}{Dan Grossman}, \bibinfo{person}{J.~Gregory
  Morrisett}, \bibinfo{person}{Trevor Jim}, \bibinfo{person}{Michael~W. Hicks},
  \bibinfo{person}{Yanling Wang}, {and} \bibinfo{person}{James Cheney}.}
  \bibinfo{year}{2002}\natexlab{}.
\newblock \showarticletitle{Region-Based Memory Management in Cyclone}. In
  \bibinfo{booktitle}{{\em Proceedings of the 2002 {ACM} {SIGPLAN} Conference
  on Programming Language Design and Implementation (PLDI), Berlin, Germany,
  June 17-19, 2002}}, \bibfield{editor}{\bibinfo{person}{Jens Knoop} {and}
  \bibinfo{person}{Laurie~J. Hendren}} (Eds.). \bibinfo{publisher}{{ACM}},
  \bibinfo{pages}{282--293}.
\newblock
\showDOI{%
\url{https://doi.org/10.1145/512529.512563}}


\bibitem[\protect\citeauthoryear{Hinnant, Dimov, and Abrahams}{Hinnant
  et~al\mbox{.}}{2002}]%
        {Hinnant2002}
\bibfield{author}{\bibinfo{person}{Howard~E. Hinnant}, \bibinfo{person}{Peter
  Dimov}, {and} \bibinfo{person}{Dave Abrahams}.}
  \bibinfo{year}{2002}\natexlab{}.
\newblock \bibinfo{title}{A Proposal to Add Move Semantics Support to the C++
  Language}.
\newblock   (\bibinfo{year}{2002}).
\newblock
\showURL{%
\url{http://www.open-std.org/jtc1/sc22/wg21/docs/papers/2002/n1377.htm}}


\bibitem[\protect\citeauthoryear{Hofmann}{Hofmann}{2000}]%
        {Hofmann2000}
\bibfield{author}{\bibinfo{person}{Martin Hofmann}.}
  \bibinfo{year}{2000}\natexlab{}.
\newblock \showarticletitle{A Type System for Bounded Space and Functional
  In-Place Update}.
\newblock \bibinfo{journal}{{\em Nord. J. Comput.\/}} \bibinfo{volume}{7},
  \bibinfo{number}{4} (\bibinfo{year}{2000}), \bibinfo{pages}{258--289}.
\newblock


\bibitem[\protect\citeauthoryear{Huet}{Huet}{1997}]%
        {Huet1997a}
\bibfield{author}{\bibinfo{person}{G{\'{e}}rard~P. Huet}.}
  \bibinfo{year}{1997}\natexlab{}.
\newblock \showarticletitle{The Zipper}.
\newblock \bibinfo{journal}{{\em J. Funct. Program.\/}} \bibinfo{volume}{7},
  \bibinfo{number}{5} (\bibinfo{year}{1997}), \bibinfo{pages}{549--554}.
\newblock
\showURL{%
\url{http://journals.cambridge.org/action/displayAbstract?aid=44121}}


\bibitem[\protect\citeauthoryear{Jim, Morrisett, Grossman, Hicks, Cheney, and
  Wang}{Jim et~al\mbox{.}}{2002}]%
        {Jim2002}
\bibfield{author}{\bibinfo{person}{Trevor Jim}, \bibinfo{person}{J.~Gregory
  Morrisett}, \bibinfo{person}{Dan Grossman}, \bibinfo{person}{Michael~W.
  Hicks}, \bibinfo{person}{James Cheney}, {and} \bibinfo{person}{Yanling
  Wang}.} \bibinfo{year}{2002}\natexlab{}.
\newblock \showarticletitle{Cyclone: {A} Safe Dialect of {C}}. In
  \bibinfo{booktitle}{{\em Proceedings of the General Track: 2002 {USENIX}
  Annual Technical Conference, June 10-15, 2002, Monterey, California, {USA}}},
  \bibfield{editor}{\bibinfo{person}{Carla~Schlatter Ellis}} (Ed.).
  \bibinfo{publisher}{{USENIX}}, \bibinfo{pages}{275--288}.
\newblock
\showURL{%
\url{http://www.usenix.org/publications/library/proceedings/usenix02/jim.html}}


\bibitem[\protect\citeauthoryear{Jung, Jourdan, Krebbers, and Dreyer}{Jung
  et~al\mbox{.}}{2018}]%
        {Jung2018}
\bibfield{author}{\bibinfo{person}{Ralf Jung}, \bibinfo{person}{Jacques{-}Henri
  Jourdan}, \bibinfo{person}{Robbert Krebbers}, {and} \bibinfo{person}{Derek
  Dreyer}.} \bibinfo{year}{2018}\natexlab{}.
\newblock \showarticletitle{RustBelt: securing the foundations of the rust
  programming language}.
\newblock \bibinfo{journal}{{\em {PACMPL}\/}} \bibinfo{volume}{2},
  \bibinfo{number}{{POPL}} (\bibinfo{year}{2018}),
  \bibinfo{pages}{66:1--66:34}.
\newblock
\showDOI{%
\url{https://doi.org/10.1145/3158154}}


\bibitem[\protect\citeauthoryear{Kobayashi}{Kobayashi}{1999}]%
        {Kobayashi1999}
\bibfield{author}{\bibinfo{person}{Naoki Kobayashi}.}
  \bibinfo{year}{1999}\natexlab{}.
\newblock \showarticletitle{Quasi-Linear Types}. In \bibinfo{booktitle}{{\em
  {POPL} '99, Proceedings of the 26th {ACM} {SIGPLAN-SIGACT} Symposium on
  Principles of Programming Languages, San Antonio, TX, USA, January 20-22,
  1999}}, \bibfield{editor}{\bibinfo{person}{Andrew~W. Appel} {and}
  \bibinfo{person}{Alex Aiken}} (Eds.). \bibinfo{publisher}{{ACM}},
  \bibinfo{pages}{29--42}.
\newblock
\showDOI{%
\url{https://doi.org/10.1145/292540.292546}}


\bibitem[\protect\citeauthoryear{Lafont}{Lafont}{1988}]%
        {Lafont1988}
\bibfield{author}{\bibinfo{person}{Yves Lafont}.}
  \bibinfo{year}{1988}\natexlab{}.
\newblock \showarticletitle{The linear abstract machine}.
\newblock \bibinfo{journal}{{\em Theoretical computer science\/}}
  \bibinfo{volume}{59}, \bibinfo{number}{1-2} (\bibinfo{year}{1988}),
  \bibinfo{pages}{157--180}.
\newblock


\bibitem[\protect\citeauthoryear{Landin}{Landin}{1965}]%
        {Landin1965J}
\bibfield{author}{\bibinfo{person}{Peter~J. Landin}.}
  \bibinfo{year}{1965}\natexlab{}.
\newblock \bibinfo{booktitle}{{\em {A} {G}eneralization of {J}umps and
  {L}abels}}.
\newblock \bibinfo{type}{{T}echnical {R}eport}.
\newblock
\newblock
\shownote{Subsequently published as~\cite{Landin1998}.}


\bibitem[\protect\citeauthoryear{Landin}{Landin}{1998}]%
        {Landin1998}
\bibfield{author}{\bibinfo{person}{Peter~J. Landin}.}
  \bibinfo{year}{1998}\natexlab{}.
\newblock \showarticletitle{{A} {G}eneralization of {J}umps and {L}abels}.
\newblock \bibinfo{journal}{{\em Higher-Order and Symbolic Computation\/}}
  \bibinfo{volume}{11}, \bibinfo{number}{2} (\bibinfo{year}{1998}),
  \bibinfo{pages}{125--143}.
\newblock


\bibitem[\protect\citeauthoryear{Leroy}{Leroy}{1990}]%
        {Leroy1990}
\bibfield{author}{\bibinfo{person}{Xavier Leroy}.}
  \bibinfo{year}{1990}\natexlab{}.
\newblock \bibinfo{booktitle}{{\em {T}he {ZINC} experiment: an economical
  implementation of the {ML} language}}.
\newblock \bibinfo{type}{{T}echnical {R}eport}. \bibinfo{institution}{INRIA}.
\newblock


\bibitem[\protect\citeauthoryear{Levy}{Levy}{1999}]%
        {Levy99CBPV}
\bibfield{author}{\bibinfo{person}{Paul~Blain Levy}.}
  \bibinfo{year}{1999}\natexlab{}.
\newblock \showarticletitle{{C}all-by-{P}ush-{V}alue: {A} {S}ubsuming
  {P}aradigm}. In \bibinfo{booktitle}{{\em Proc. TLCA '99}}.
  \bibinfo{pages}{228--242}.
\newblock


\bibitem[\protect\citeauthoryear{Levy}{Levy}{2004}]%
        {Levy2004}
\bibfield{author}{\bibinfo{person}{Paul~Blain Levy}.}
  \bibinfo{year}{2004}\natexlab{}.
\newblock \bibinfo{booktitle}{{\em {C}all-{B}y-{P}ush-{V}alue: {A}
  {F}unctional/{I}mperative {S}ynthesis}}. \bibinfo{series}{Semantic Structures
  in Computation}, Vol.~\bibinfo{volume}{2}.
\newblock \bibinfo{publisher}{Springer}.
\newblock
\showISBNx{1-4020-1730-8}


\bibitem[\protect\citeauthoryear{Melli\`{e}s}{Melli\`{e}s}{2009}]%
        {PAM2009}
\bibfield{author}{\bibinfo{person}{Paul-Andr\'{e} Melli\`{e}s}.}
  \bibinfo{year}{2009}\natexlab{}.
\newblock \bibinfo{booktitle}{{\em {C}ategorical semantics of linear logic}}.
  \bibinfo{series}{Panoramas et Synth\`{e}ses}, Vol.~\bibinfo{volume}{27}.
\newblock \bibinfo{publisher}{Soci\'{e}t\'{e} Math\'{e}matique de France},
  Chapter~1, \bibinfo{pages}{15--215}.
\newblock


\bibitem[\protect\citeauthoryear{Melli{\`e}s and Tabareau}{Melli{\`e}s and
  Tabareau}{2010}]%
        {MelTab10RessourceModalities}
\bibfield{author}{\bibinfo{person}{Paul-Andr{\'e} Melli{\`e}s} {and}
  \bibinfo{person}{Nicolas Tabareau}.} \bibinfo{year}{2010}\natexlab{}.
\newblock \showarticletitle{{R}esource modalities in tensor logic}.
\newblock \bibinfo{journal}{{\em Ann. Pure Appl. Logic\/}}
  \bibinfo{volume}{161}, \bibinfo{number}{5} (\bibinfo{year}{2010}),
  \bibinfo{pages}{632--653}.
\newblock


\bibitem[\protect\citeauthoryear{Minsky}{Minsky}{1996}]%
        {Minsky1996b}
\bibfield{author}{\bibinfo{person}{Naftaly~H. Minsky}.}
  \bibinfo{year}{1996}\natexlab{}.
\newblock \showarticletitle{Towards Alias-Free Pointers}. In
  \bibinfo{booktitle}{{\em ECOOP'96 - Object-Oriented Programming, 10th
  European Conference, Linz, Austria, July 8-12, 1996, Proceedings}} {\em
  (\bibinfo{series}{Lecture Notes in Computer Science})},
  \bibfield{editor}{\bibinfo{person}{Pierre Cointe}} (Ed.),
  Vol.~\bibinfo{volume}{1098}. \bibinfo{publisher}{Springer},
  \bibinfo{pages}{189--209}.
\newblock
\showDOI{%
\url{https://doi.org/10.1007/BFb0053062}}


\bibitem[\protect\citeauthoryear{Minsky, Madhavapeddy, and Hickey}{Minsky
  et~al\mbox{.}}{2013}]%
        {Minsky2013}
\bibfield{author}{\bibinfo{person}{Yaron Minsky}, \bibinfo{person}{Anil
  Madhavapeddy}, {and} \bibinfo{person}{Jason Hickey}.}
  \bibinfo{year}{2013}\natexlab{}.
\newblock \bibinfo{booktitle}{{\em Real World OCaml - Functional Programming
  for the Masses}}.
\newblock \bibinfo{publisher}{O'Reilly}.
\newblock
\showISBNx{978-1-4493-2391-2}


\bibitem[\protect\citeauthoryear{Morrisett}{Morrisett}{1995}]%
        {Morrisett1995}
\bibfield{author}{\bibinfo{person}{Greg Morrisett}.}
  \bibinfo{year}{1995}\natexlab{}.
\newblock \bibinfo{title}{Compiling with Types}.
\newblock   (\bibinfo{year}{1995}).
\newblock


\bibitem[\protect\citeauthoryear{Munch-Maccagnoni}{Munch-Maccagnoni}{2014}]%
        {Munch14Involutive}
\bibfield{author}{\bibinfo{person}{Guillaume Munch-Maccagnoni}.}
  \bibinfo{year}{2014}\natexlab{}.
\newblock \showarticletitle{{F}ormulae-as-{T}ypes for an {I}nvolutive
  {N}egation}. In \bibinfo{booktitle}{{\em Proceedings of the joint meeting of
  the Twenty-Third EACSL Annual Conference on Computer Science Logic and the
  Twenty-Ninth Annual ACM/IEEE Symposium on Logic in Computer Science
  (CSL-LICS)}}.
\newblock


\bibitem[\protect\citeauthoryear{O'Hearn, Power, Takeyama, and Tennent}{O'Hearn
  et~al\mbox{.}}{1999}]%
        {OHearn1999}
\bibfield{author}{\bibinfo{person}{Peter~W. O'Hearn}, \bibinfo{person}{John
  Power}, \bibinfo{person}{Makoto Takeyama}, {and} \bibinfo{person}{Robert~D.
  Tennent}.} \bibinfo{year}{1999}\natexlab{}.
\newblock \showarticletitle{Syntactic Control of Interference Revisited}.
\newblock \bibinfo{journal}{{\em Theor. Comput. Sci.\/}} \bibinfo{volume}{228},
  \bibinfo{number}{1-2} (\bibinfo{year}{1999}), \bibinfo{pages}{211--252}.
\newblock
\showDOI{%
\url{https://doi.org/10.1016/S0304-3975(98)00359-4}}


\bibitem[\protect\citeauthoryear{Pottier and Protzenko}{Pottier and
  Protzenko}{2013}]%
        {Pottier2013}
\bibfield{author}{\bibinfo{person}{Fran{\c{c}}ois Pottier} {and}
  \bibinfo{person}{Jonathan Protzenko}.} \bibinfo{year}{2013}\natexlab{}.
\newblock \showarticletitle{Programming with permissions in Mezzo}. In
  \bibinfo{booktitle}{{\em ACM SIGPLAN Notices}}, Vol.~\bibinfo{volume}{48}.
  ACM, \bibinfo{pages}{173--184}.
\newblock


\bibitem[\protect\citeauthoryear{Proust}{Proust}{2016}]%
        {Proust2016}
\bibfield{author}{\bibinfo{person}{Rapha{\"e}l Proust}.}
  \bibinfo{year}{2016}\natexlab{}.
\newblock \bibinfo{title}{ASAP: As Static as Possible memory management}.
\newblock   (\bibinfo{year}{2016}).
\newblock
\showURL{%
\url{http://www.cl.cam.ac.uk/techreports/UCAM-CL-TR-908.pdf}}


\bibitem[\protect\citeauthoryear{Ramananandro, Dos~Reis, and
  Leroy}{Ramananandro et~al\mbox{.}}{2012}]%
        {Ramananandro2012}
\bibfield{author}{\bibinfo{person}{Tahina Ramananandro},
  \bibinfo{person}{Gabriel Dos~Reis}, {and} \bibinfo{person}{Xavier Leroy}.}
  \bibinfo{year}{2012}\natexlab{}.
\newblock \showarticletitle{A mechanized semantics for C++ object construction
  and destruction, with applications to resource management}. In
  \bibinfo{booktitle}{{\em ACM SIGPLAN Notices}}, Vol.~\bibinfo{volume}{47}.
  ACM, \bibinfo{pages}{521--532}.
\newblock


\bibitem[\protect\citeauthoryear{Reynolds}{Reynolds}{1978}]%
        {Reynolds1978}
\bibfield{author}{\bibinfo{person}{John~C. Reynolds}.}
  \bibinfo{year}{1978}\natexlab{}.
\newblock \showarticletitle{Syntactic Control of Interference}. In
  \bibinfo{booktitle}{{\em Conference Record of the Fifth Annual {ACM}
  Symposium on Principles of Programming Languages, Tucson, Arizona, USA,
  January 1978}}, \bibfield{editor}{\bibinfo{person}{Alfred~V. Aho},
  \bibinfo{person}{Stephen~N. Zilles}, {and} \bibinfo{person}{Thomas~G.
  Szymanski}} (Eds.). \bibinfo{publisher}{{ACM} Press},
  \bibinfo{pages}{39--46}.
\newblock
\showDOI{%
\url{https://doi.org/10.1145/512760.512766}}


\bibitem[\protect\citeauthoryear{Shi and Xi}{Shi and Xi}{2013}]%
        {Shi2013}
\bibfield{author}{\bibinfo{person}{Rui Shi} {and} \bibinfo{person}{Hongwei
  Xi}.} \bibinfo{year}{2013}\natexlab{}.
\newblock \showarticletitle{A linear type system for multicore programming in
  {ATS}}.
\newblock \bibinfo{journal}{{\em Sci. Comput. Program.\/}}
  \bibinfo{volume}{78}, \bibinfo{number}{8} (\bibinfo{year}{2013}),
  \bibinfo{pages}{1176--1192}.
\newblock
\showDOI{%
\url{https://doi.org/10.1016/j.scico.2012.09.005}}


\bibitem[\protect\citeauthoryear{Spiwack}{Spiwack}{2014}]%
        {Spiwak2014}
\bibfield{author}{\bibinfo{person}{Arnaud Spiwack}.}
  \bibinfo{year}{2014}\natexlab{}.
\newblock \bibinfo{title}{{A} dissection of {L}}.  (\bibinfo{year}{2014}).
\newblock
\newblock
\shownote{Manuscript.}


\bibitem[\protect\citeauthoryear{Stroustrup}{Stroustrup}{1994}]%
        {Stroustrup1994}
\bibfield{author}{\bibinfo{person}{Bjarne Stroustrup}.}
  \bibinfo{year}{1994}\natexlab{}.
\newblock \bibinfo{booktitle}{{\em The design and evolution of {C}++}}.
\newblock \bibinfo{publisher}{Pearson Education India}.
\newblock


\bibitem[\protect\citeauthoryear{Stroustrup}{Stroustrup}{2001}]%
        {Stroustrup2001}
\bibfield{author}{\bibinfo{person}{Bjarne Stroustrup}.}
  \bibinfo{year}{2001}\natexlab{}.
\newblock \bibinfo{booktitle}{{\em Exception Safety: Concepts and Techniques}}.
\newblock \bibinfo{publisher}{Springer Berlin Heidelberg},
  \bibinfo{address}{Berlin, Heidelberg}, \bibinfo{pages}{60--76}.
\newblock
\showISBNx{978-3-540-45407-6}
\showDOI{%
\url{https://doi.org/10.1007/3-540-45407-1_4}}


\bibitem[\protect\citeauthoryear{Stroustrup and Sutter}{Stroustrup and
  Sutter}{2015}]%
        {Stroustrup2015Core}
\bibfield{author}{\bibinfo{person}{Bjarne Stroustrup} {and}
  \bibinfo{person}{Herb Sutter}.} \bibinfo{year}{2015}\natexlab{}.
\newblock \bibinfo{title}{C++ Core Guidelines}.
\newblock   (\bibinfo{year}{2015}).
\newblock
\showURL{%
\url{https://github.com/isocpp/CppCoreGuidelines}}


\bibitem[\protect\citeauthoryear{Stroustrup, Sutter, and Dos~Reis}{Stroustrup
  et~al\mbox{.}}{2015}]%
        {Stroustrup2015}
\bibfield{author}{\bibinfo{person}{Bjarne Stroustrup}, \bibinfo{person}{Herb
  Sutter}, {and} \bibinfo{person}{Gabriel Dos~Reis}.}
  \bibinfo{year}{2015}\natexlab{}.
\newblock \bibinfo{title}{A brief introduction to {C}++'s model for type- and
  resource-safety}.
\newblock   (\bibinfo{year}{2015}).
\newblock
\showURL{%
\url{https://github.com/isocpp/CppCoreGuidelines/blob/master/docs/Introductiontotypeandresourcesafety.pdf}}


\bibitem[\protect\citeauthoryear{Tofte and Birkedal}{Tofte and
  Birkedal}{1998}]%
        {Tofte1998}
\bibfield{author}{\bibinfo{person}{Mads Tofte} {and} \bibinfo{person}{Lars
  Birkedal}.} \bibinfo{year}{1998}\natexlab{}.
\newblock \showarticletitle{A region inference algorithm}.
\newblock \bibinfo{journal}{{\em ACM Transactions on Programming Languages and
  Systems (TOPLAS)\/}} \bibinfo{volume}{20}, \bibinfo{number}{4}
  (\bibinfo{year}{1998}), \bibinfo{pages}{724--767}.
\newblock


\bibitem[\protect\citeauthoryear{Tofte and Talpin}{Tofte and Talpin}{1994}]%
        {Tofte1994}
\bibfield{author}{\bibinfo{person}{Mads Tofte} {and}
  \bibinfo{person}{Jean{-}Pierre Talpin}.} \bibinfo{year}{1994}\natexlab{}.
\newblock \showarticletitle{Implementation of the Typed Call-by-Value
  lambda-Calculus using a Stack of Regions}. In \bibinfo{booktitle}{{\em
  Conference Record of POPL'94: 21st {ACM} {SIGPLAN-SIGACT} Symposium on
  Principles of Programming Languages, Portland, Oregon, USA, January 17-21,
  1994}}, \bibfield{editor}{\bibinfo{person}{Hans{-}Juergen Boehm},
  \bibinfo{person}{Bernard Lang}, {and} \bibinfo{person}{Daniel~M. Yellin}}
  (Eds.). \bibinfo{publisher}{{ACM} Press}, \bibinfo{pages}{188--201}.
\newblock
\showDOI{%
\url{https://doi.org/10.1145/174675.177855}}


\bibitem[\protect\citeauthoryear{Tov and Pucella}{Tov and Pucella}{2011}]%
        {Tov2011}
\bibfield{author}{\bibinfo{person}{Jesse~A. Tov} {and}
  \bibinfo{person}{Riccardo Pucella}.} \bibinfo{year}{2011}\natexlab{}.
\newblock \showarticletitle{Practical affine types}. In
  \bibinfo{booktitle}{{\em Proceedings of the 38th {ACM} {SIGPLAN-SIGACT}
  Symposium on Principles of Programming Languages, {POPL} 2011, Austin, TX,
  USA, January 26-28, 2011}}, \bibfield{editor}{\bibinfo{person}{Thomas Ball}
  {and} \bibinfo{person}{Mooly Sagiv}} (Eds.). \bibinfo{publisher}{{ACM}},
  \bibinfo{pages}{447--458}.
\newblock
\showDOI{%
\url{https://doi.org/10.1145/1926385.1926436}}


\bibitem[\protect\citeauthoryear{Wadler}{Wadler}{1990}]%
        {Wadler90lineartypes}
\bibfield{author}{\bibinfo{person}{Philip Wadler}.}
  \bibinfo{year}{1990}\natexlab{}.
\newblock \showarticletitle{Linear Types Can Change the World!}. In
  \bibinfo{booktitle}{{\em Programming Concepts and Methods}}.
  \bibinfo{publisher}{North}.
\newblock


\bibitem[\protect\citeauthoryear{Walker}{Walker}{2005}]%
        {Walker2005a}
\bibfield{author}{\bibinfo{person}{David Walker}.}
  \bibinfo{year}{2005}\natexlab{}.
\newblock \showarticletitle{Substructural type systems}.
\newblock In \bibinfo{booktitle}{{\em Advanced Topics in Types and Programming
  Languages}}, \bibfield{editor}{\bibinfo{person}{Benjamin~C. Pierce}} (Ed.).
  \bibinfo{publisher}{The MIT Press}, \bibinfo{pages}{3--44}.
\newblock


\bibitem[\protect\citeauthoryear{Zhu and Xi}{Zhu and Xi}{2005}]%
        {Zhu2005}
\bibfield{author}{\bibinfo{person}{Dengping Zhu} {and} \bibinfo{person}{Hongwei
  Xi}.} \bibinfo{year}{2005}\natexlab{}.
\newblock \showarticletitle{Safe Programming with Pointers Through Stateful
  Views}. In \bibinfo{booktitle}{{\em Practical Aspects of Declarative
  Languages, 7th International Symposium, {PADL} 2005, Long Beach, CA, USA,
  January 10-11, 2005, Proceedings}} {\em (\bibinfo{series}{Lecture Notes in
  Computer Science})}, \bibfield{editor}{\bibinfo{person}{Manuel~V.
  Hermenegildo} {and} \bibinfo{person}{Daniel Cabeza}} (Eds.),
  Vol.~\bibinfo{volume}{3350}. \bibinfo{publisher}{Springer},
  \bibinfo{pages}{83--97}.
\newblock
\showDOI{%
\url{https://doi.org/10.1007/978-3-540-30557-6_8}}


\end{thebibliography}

\end{document}